\documentclass[twocolumn]{aastex631}

\usepackage{amsmath}
\usepackage{bm}
\usepackage{hyperref}
\usepackage{enumitem}
\usepackage{xcolor}
\usepackage{soul}
\usepackage{graphicx}
\graphicspath{{figures_lr/}}

\usepackage{ulem} 

\newcommand{\rmscriptsize}[1]{\textrm{\scriptsize{#1}}}
\newcommand{\SUB}[3][]{
  \ifx\relax#1\relax
  \ensuremath{#2_{\rmscriptsize{#3}}}
  \else
  \ensuremath{#2_{#3,\,\rmscriptsize{#1}}}
  \fi}

\newcommand{\UP}[3][]{
  \ifx\relax#1\relax
  \ensuremath{#2^{\rmscriptsize{#3}}}
  \else
  \ensuremath{#2^{#3,\,\rmscriptsize{#1}}}
  \fi}
\newcommand{\GJ}[1]{\SUB{#1}{\textsc{gj}}}
\newcommand{\UGJ}[1]{\UP{#1}{\textsc{gj}}}
\newcommand{\PC}[1]{\SUB{#1}{\textsc{pc}}}
\newcommand{\LC}[1]{\SUB{#1}{\textsc{lc}}}
\newcommand{\FFE}[1]{\SUB{#1}{\textsc{ffe}}}
\newcommand{\fine}[1]{\SUB{#1}{\textsc{f}}}

\definecolor{light-gray}{gray}{0.8}
\newcommand{\V}[1]{\mathbf{#1}}

\shorttitle{Coherence scale in pulsar discharge}
\shortauthors{Chernoglazov, A. et al.}

\begin{document}

\title{Coherence of Multi-Dimensional Pair Production Discharges in Polar Caps of Pulsars}

\correspondingauthor{Alexander Chernoglazov}
\email{achernog@ias.edu}

\author[0000-0001-5121-1594]{Alexander Chernoglazov}
\affiliation{Department of Physics, University of Maryland, College Park, MD 20742, USA}
\affiliation{Institute for Research in Electronics and Applied Physics, University of Maryland, College Park, MD 20742, USA}
\affiliation{Institute for Advanced Study, Princeton, NJ 08540, USA}

\author[0000-0001-7801-0362]{Alexander Philippov}
\affiliation{Department of Physics, University of Maryland, College Park, MD 20742, USA}
\affiliation{Institute for Research in Electronics and Applied Physics, University of Maryland, College Park, MD 20742, USA}

\author[0000-0002-0067-1272]{Andrey Timokhin}
\affiliation{Janusz Gil Institute of Astronomy, University of Zielona G\'{o}ra, ul. Szafrana 2, 65516, Zielona G\'{o}ra, Poland}

\begin{abstract}

We report on the first self-consistent multidimensional particle-in-cell numerical simulations of non-homogeneous pair discharges in polar caps of rotation-powered pulsars. By introducing strong inhomogeneities in the initial plasma distribution in our simulations, we analyze the degree of self-consistently emerging coherence of discharges across magnetic field lines. In 2D, we study discharge evolution for a wide range of physical parameters and boundary conditions corresponding to both the absent and free escape of charged particles from the surface of a neutron star. We also present the results of the first 3D simulations of discharges in a polar cap with a distribution of the global magnetospheric current appropriate for a pulsar with $60^{\circ}$ inclination angle. For all parameters, we find the coherence scale of pair discharges across magnetic field lines to be of the order of gap height. We also demonstrate that the popular “spark” model of pair discharges is incompatible with the universally adopted force-free magnetosphere model: intermittent discharges fill the entire zone of the polar cap that allows pair cascades, leaving no space for discharge-free regions. Our findings disprove the key assumption of the spark model about the existence of isolated distinct discharge columns. 

\end{abstract}


\section{Introduction} \label{sec:intro}

Pulsars are highly magnetized (up to $\sim{}10^{13}$G) rapidly-spinning neutron stars (NS) \citep[e.g.,][]{Philippov2022}. They emit pulsed non-thermal radiation across a wide range of frequencies generated by dense pair plasma in their magnetospheres. Most of the plasma is produced in quantum-electrodynamic (QED) discharges in polar caps -- regions around magnetic poles close to the NS surface. There, a rotation-induced electric field pulls electrons from the atmosphere of the NS and accelerates them to high energies. As these electrons propagate along the curved magnetic field lines, they emit high-energy curvature photons, which subsequently decay into electron-positron pairs as first suggested by \citet{Sturrock1971}. For a long time, the de facto standard model for pair production assumed a stationary unidirectional space-charge-limited flow (SCLF) of plasma extracted from the NS atmosphere. Here, electrons are accelerated by the electric field due to charge starvation caused by a small mismatch between the local Goldreich-Julian (GJ, \citealt{Goldreich1969}) and the actual plasma charge densities, $\rho-\GJ{\rho}$ \citep{Arons1979, Muslimov1992}. However, it was later demonstrated that the current density needed to support the twist of open magnetic field lines in the pulsar magnetosphere, $j_{\rm mag}$, could not be sustained by a unidirectional plasma flow \citep{Timokhin2006}, in general requiring a flow with counter-streaming particles. In such flows, charge starvation, and thus the accelerating electric field, appears when there is a difference between the GJ current density, $\GJ{j}=\GJ{\rho}c$, and the current density required by the magnetosphere, $j_{\rm mag}-\GJ{j}$; thus, the accelerating electric field is ``current-density-driven'', rather than ``charge-density-driven'' \citep{Beloborodov2008, Timokhin2010, Timokhin2013}.  Self-consistent particle-in-cell (PIC) plasma simulations \citep{Timokhin2010, Timokhin2013}, which took into account particle acceleration, photon emission, and pair production, demonstrated that, regardless of the ability of particles to escape from the NS surface, pair discharges always operate in a cyclic fashion. These findings are reminiscent of the model of \citet{Ruderman1975} (RS) with no particle outflow from the NS surface. The bursts of pair formation are followed by a quiet outflow of plasma where the electric field is screened; once a substantial amount of the produced plasma leaves the polar cap, the electric field is restored, and the cycle restarts. 

For most pulsars, the radio emission is generated in the inner magnetosphere not far from their polar caps \citep[e.g.,][]{Lyne2004, Philippov2022}. The observed brightness temperature of the radio emission is incredibly high, $10^{25}-10^{30}$ K \citep{Lorimer2012}, indicating a coherent nature of the radiation mechanism. Most of the existing models of pulsar radio emission assume the development of plasma instabilities in the pair plasma above the polar cap when the accelerating electric field is already screened \citep{Melrose2021}. Recently, \cite{Philippov2020} proposed a novel radio emission mechanism driven by the screening of the accelerating electric field during pair discharge. Their 2D PIC simulations demonstrated the excitation of electromagnetic superluminal modes when the discharge occurs obliquely with respect to the background magnetic field, i.e., in this model, the radio emission is caused by the spatial inhomogeneity of the pair discharges. Several studies have confirmed the excitation of electromagnetic waves through \cite{Philippov2020} mechanism using 2D local \citep{Cruz201, Benavek2024} and global \citep{Bransgrove2023} PIC simulations. However, the scale at which the discharge maintains spatial coherence across magnetic field lines has not yet been explored, partially because of the limited separation of plasma and global scales and the simplified physics of pair production in multi-dimensional simulations. Spatial decorrelation of the polar discharge across the field lines might seem plausible as a physics basis to explain some aspects of the short-term nulling of the radio emission \citep[e.g.,][]{Ng2020}.

Observed radio emission shows variability over a wide range of timescales. This variability includes the microstructure of individual radio pulses at a level of $\sim 100$  microseconds and smaller, as well as the quasi-stable drift of individual sub-pulses seen in some pulsars \citep[e.g.,][]{Drake1968, Janagal2023}. To explain these phenomena, RS proposed that pair discharges occur in a set of localized columns, which they called ``sparks", separated by plasma-depleted regions. The potential drop in the sparks gives rise to the electric field component perpendicular to the magnetic field, which causes sparks' quasi-regular drift with respect to the NS. The spark model is still being used for modeling drifting subpulses \citep[e.g.][]{GilSendyk2000, GilMelikidzeGeppert2003, BasuMitraMelikidze2020}. The intermittency of the cyclic discharge, as revealed by simulations, is a natural candidate for explaining the microstructure. On the other hand, the existence of long-lived isolated sparks in intermittent discharges is questionable. 

In this Letter, we aim to explore the existence and sustainability of the spatial de-correlations in the pair discharges for the polar cap models with no \citep{Ruderman1975} as well as free \citep{Arons1979} escape of particles from the NS surface. Using 2D and 3D PIC simulations, we study how pair discharges fill the polar cap and the characteristic discharge scales perpendicular to the magnetic field. Our paper is organized as follows. In Section \ref{overview}, we describe the physical model employed in our simulations of polar cap discharges. In Section \ref{results}, we present the main results of our simulations. We demonstrate that the discharge desynchronizes on transverse scales comparable to the length of the gap along the field line and conclude that localized sparks can not exist in polar caps of pulsars with force-free magnetospheres. Section \ref{discussion} discusses the applications of our results to observed pulsars.

\section{Overview of the physical model}
\label{overview}

\subsection{Dynamics of the electromagnetic fields}
\label{GapDynamics}
Global solutions for the pulsar magnetospheres in the force-free (FFE) approximation \citep[e.g.,][]{Contopoulos1999, Spitkovsky2006, Timokhin2006,  Kalapotharakos2012} demonstrate that the open field lines develop a twist that is self-regulated by conditions at the light cylinder. The global toroidal magnetic field, ${\bm B}_\varphi$, sets the distribution of the field-aligned magnetospheric current, \mbox{${\bm j}_{\rm mag} \equiv (c/4\pi) \nabla\times {\bm B}_\varphi$}, in the polar cap zone.  We decompose the electromagnetic fields into a sum of the FFE fields, ${\FFE{\bm B}={{\bm B}_0}+{\bm B}_\varphi}$, $\FFE{{\bm E}}=-{\bm \Omega}\times {\bm{r}} \times {{\bm B}_0}/c$, where ${{\bm B}_0}$ is the background (e.g., dipolar) poloidal magnetic field, and corrections, $\delta{\bm{E}}$ and $\delta{\bm{B}}$, created in the process of adjustment of the electric current in the gap, ${\bm j}$, to the global magnetospheric current, \mbox{${\bm j}_{\rm mag}$}. By substituting the field decomposition into time-dependent Maxwell equations, we obtain the evolution equations for $\delta\bm{E}$ and $ \delta {\bm B}$: 
\begin{eqnarray}
    \frac{\partial }{\partial t} \delta {\bm E} &=& c \nabla\times\delta {\bm B} - 4 \pi ({\bm j} - {\bm j}_{\rm mag}), \label{FieldEvolution1}\\
    \frac{\partial }{\partial t} \delta \bm{B}&=& - c\nabla\times \delta {\bm E}.
\label{FieldEvolution2}
\end{eqnarray}
The electric field has also to obey Gauss's law:
\begin{equation}
\nabla \cdot  \delta {\bm E} = 4 \pi (\rho-\GJ{\rho}), 
\label{FieldConstrain}
\end{equation}
where $\rho$ is the charge density, and $\GJ{\rho} = \nabla \cdot \FFE{\bm E}/4\pi = -{\bm \Omega}\cdot {\bm B}/2\pi c$ is the GJ charge density. Unless mentioned otherwise, we consider the GJ density to be negative.

We assume that the distribution of magnetospheric current, \mbox{${\bm j}_{\rm mag}$}, is constant in our simulations. This is well justified in actively pair-producing pulsars, in which case the global toroidal field, ${\bm B}_\varphi$, evolves on timescales comparable to the light crossing time of the light cylinder, $\LC{R}/c$, which is much longer than the characteristic time in the polar cap zone, $\PC{R}/c$. Here, $\PC{R}=R_\star\sqrt{R_\star/\LC{R}}$ is the radius of the polar cap, $\LC{R}=c P/2\pi$ is the radius of the light cylinder, $R_\star$ is the stellar radius, and $P$ is the rotational period of the star. Equations (\ref{FieldEvolution1}) and (\ref{FieldEvolution2}) have two special stationary, $\partial/\partial t = 0$, solutions:
\begin{enumerate}[label=(\roman*)]
    \item There is abundant plasma everywhere to support the magnetospheric current, $j = j_{\rm mag}$, which leads to a fully FFE solution, $\delta {\bm E}, \delta {\bm B} = 0$. 
    \item A field line has no plasma loading and $j=0$ in the gap zone. Then, the full twist of the field lines becomes zero, $\delta B = - B_\varphi$.
\end{enumerate}

In the previously studied 1D approximation, the set of Eqns.(\ref{FieldEvolution1}-\ref{FieldEvolution2}) reduces to the $\partial (\delta E_{||})/\partial t = - 4\pi(j_{||}-j_{\rm mag})$, since only components of the field and current parallel to the background field can be evolved in 1D \citep{Beloborodov2008, Timokhin2010, Timokhin2013}. By construction, these solutions describe deviations from the FFE state, i.e., solution (i). Our multidimensional simulations solve the full system of Eqns. (\ref{FieldEvolution1}-\ref{FieldEvolution2}) together with the constraint Eqn. (\ref{FieldConstrain}). In contrast to the 1D simulations, if the pair creation process is unable to sustain the global magnetospheric current, the total twist can become zero, and the system reaches state (ii).  

The limitations of numerical methods do not allow the study of processes in pulsar polar caps for realistic values of physical parameters except in 1D. One must rely on a scaled model with sometimes exaggerated parameters to make such a study possible in 2D and 3D. Despite the simplifications, our numerical setup preserves all the important properties of the real polar cap at the base of the FFE magnetosphere, as described below.

\subsection{Magnetospheric current distribution}

\begin{figure}[t]
\includegraphics[width=\columnwidth]{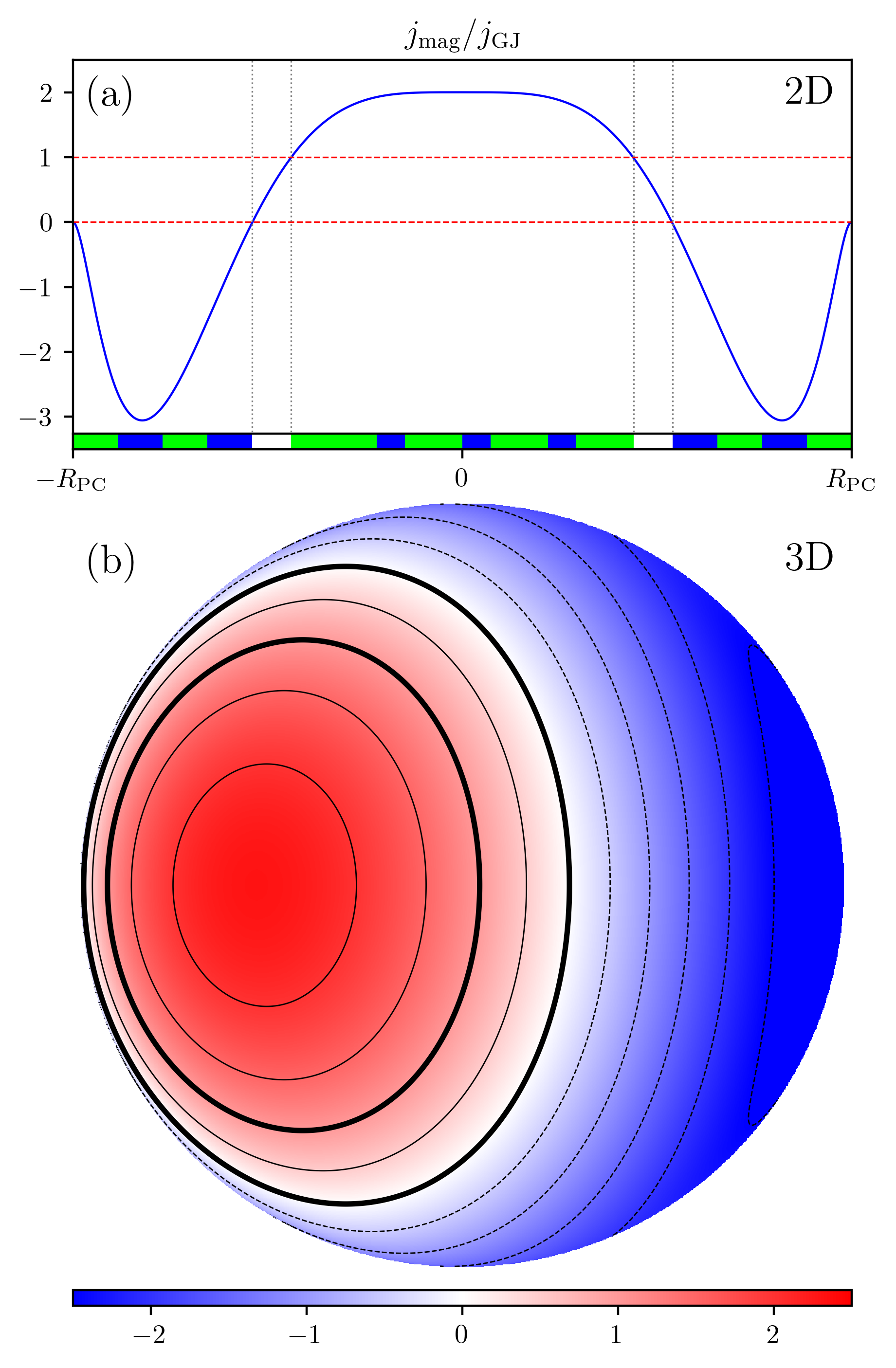}
\caption{Distribution of the magnetospheric current across the polar cap used in our (a) 2D and (b) 3D simulations, which mimic the currents in the magnetospheres of an aligned and $60^\circ$-inclined rotators. The net current across the polar cap is zero, with outgoing and return current regions each covering roughly half of the polar cap. The colored inset in panel (a) marks the boundaries of initial high-density plasma-filled patches (see Sec.~\ref{sim:initial}). In panel (b), the thick black lines represent the boundaries of super GJ and return current regions.}
\label{fig:current}
\end{figure}

The work by \cite{Gralla2016, Gralla2017} provided fits of the spatial distribution of the magnetospheric current over the polar cap of oblique rotators. They found that for misaligned rotators, a significant fraction of the polar cap is capable of supporting active pair production, i.e., possesses super-GJ current, $j_{\rm mag}/\GJ{j} > 1$, where again $\GJ{j} = \GJ{\rho} c$, or the volume return current, $j_{\rm mag}/\GJ{j} < 0$ \citep{Timokhin2013}. 

For a 2D simulation performed in Cartesian coordinates, we choose the profile of the magnetospheric current to be qualitatively similar to the solution for the aligned rotator: 
\begin{equation}
    j_{\rm mag}^{\rm 2D}(x) = 2 - C_1 x^4 + C_2 x^6 + C_3 x^{40},
\end{equation}
here $x=r_\perp/\PC{R}$ is the coordinate transverse to the background magnetic field, $B_0$, which is chosen to be uniform.  Coefficients $C_1, C_2, C_3$ are chosen to satisfy the conditions of total zero current and smooth transition to the zone with no current (closed field lines zone):
\begin{equation}
    \int_0^{\PC{R}} j_{\rm mag}^{\rm 2D} d x =0,~~j_{\rm mag}^{\rm 2D}(1)=0,~~ \frac{d}{d x} j_{\rm mag}^{\rm 2D}\bigg|_{x=1}= 0.
\end{equation}
This choice results in approximately half of the polar cap field lines having the volume return magnetospheric current, $j_{\rm mag}<0$, and half of the field lines carry a super-GJ current, $j_{\rm mag}>\GJ{j}$\footnote{Our current density distribution has a higher maximum value and results in larger portions of the polar cap occupied by the super-GJ current density compared to a real axisymmetric pulsar. However, qualitatively, it preserves the properties of an actual current density distribution.}. The current profile is visualized in Fig. \ref{fig:current}a. Our choice excludes the extremely narrow zone of the intense return current at the boundary of the open field line region, the ``separatrix'' current, which is present in magnetospheric solutions for any obliquity. Studying the discharge in the separatrix requires fully global simulations \citep[e.g.,][]{Philippov2015,Hu2022}.

In 3D simulations, the approximation for the distribution of the magnetospheric current over the polar cap is taken from \cite{Gralla2016}:
\begin{eqnarray}
    \frac{j_{\rm mag}^{\rm 3D}}{\GJ{\rho} c} (\theta, \phi) &\approx& \frac{1}{(1-\Omega_Z/\Omega)} [J_0\left( \arcsin{(\theta/{\sqrt{\alpha_0}})}\right)-\nonumber \\ 
    & &-J_1\left(\arcsin{(\theta/{\sqrt{\alpha_0}})}\right)\tan{i}\cos{\phi}].
    \label{3dcurrent}
\end{eqnarray}
Here, $J_{0,1}$ are Bessel functions of the zeroth and first order, ${\theta}=r_\perp/R_{\star}$ gives the distance from the center of the polar cap, and $\phi$ is an azimuthal angle. The constant parameters $i,~\Omega_Z,~\alpha_0$ describe the global magnetosphere, namely $i=60^\circ$ is an inclination angle, $\Omega_Z \ \approx (2C/5) \Omega$ is the angular velocity of the frame dragging, and $C=2M/R_\star\approx0.5$ is stellar compactness, and $\alpha_0$ defines the size of the polar cap for the dipolar magnetic field. The analytical formula (\ref{3dcurrent}) has a non-zero total current through the polar cap, requiring a separatrix current. In our numerical setup, the mean current $\PC{\langle j_{\rm mag}\rangle}$ is subtracted from (\ref{3dcurrent}) to enforce the zero total volumetric current. The resulting distribution of the magnetospheric current is shown in Fig.\ref{fig:current}b. 

\begin{table*}
\caption{\label{tab:sim}Summary of simulation parameters. The simulations of 2D SCLF polar cap are discussed in section \ref{sub:sclf}. Namely, simulation \texttt{SCLFsmall} is described in subsection \ref{sec:SCLFsmall}, \texttt{SCLFsmallTail} - in \ref{sec:SCLFtails}, \texttt{SCLFsmallCurv} - in \ref{sec:SCLFdipole}, and \texttt{SCLFlarge} - in subsection \ref{sec:SCLFlarge}. Both simulations of the 2D RS model of the polar cap are presented in section \ref{sub:RS}, and the 3D simulation of the SCLF model is discussed in section \ref{sec:3D}.}
\begin{tabular}{lccccccc}
\hline \hline
\textrm{Simulation}&
\textrm{$l_{\rm gap}/\PC{R}$}&
\textrm{$\PC{R}/d_{\rm e}^{\rm GJ}$}&
\textrm{$\PC{\gamma}$}&
\textrm{$\gamma_{\rm rad}$}&
\textrm{$\gamma_{\rm emit}$}&
\textrm{$\rho_{{\rm c},0}/\PC{R}$}&
\textrm{$B_0/B_{q}$}\\
\hline
 \texttt{SCLFsmall} & $l_{\rm gap} \ll \PC{R}$ & 12000 & $7.2\times10^7$ & $6.7 \times10^5$ & $10^4$ & 13 & 2.0\\
 \texttt{SCLFsmallTail} & $l_{\rm gap} \ll \PC{R}$ & 12000 & $7.2\times10^7$ & $1.35 \times10^5$ & $4 \times 10^3$ & 13 & 2.0\\
 \texttt{SCLFsmallCurv} & $l_{\rm gap} \ll \PC{R}$ & 12000 & $7.2\times10^7$ & $6.7 \times10^5$ & $10^4$ & 13 & 2.0\\
 \texttt{SCLFlarge} & $l_{\rm gap} \lesssim \PC{R}$ & 8000 & $3.2\times10^7$ & $8 \times10^5$ & $3\times10^4$ & 20 & 0.1\\
 \hline 
 \texttt{RSsmall} & $l_{\rm gap} \ll \PC{R}$ & 8000 & $3.2\times10^7$ & $8 \times10^5$ & $3\times10^4$ & 20 & 1.0\\
 \texttt{RSlarge} & $l_{\rm gap} \lesssim \PC{R}$ & 8000 & $3.2\times10^7$ & $8 \times10^5$ & $7\times10^4$ & 20 & 0.1\\
 \hline
 \texttt{SCLF3D} & $l_{\rm gap} \lesssim \PC{R}$ & 1000 & $5\times10^5$ & $10^4$ & $10^3$ & 5 & 0.2\\
 \hline \hline      
\end{tabular}
\end{table*} 

\subsection{QED pair production}
1D studies by \cite{Timokhin2010, Timokhin2013} showed that pair production in polar zones operates intermittently, i.e., episodes with strong electric field component along the magnetic field and intense particle acceleration are followed by plasma-filled states with fully screened electric fields. During the active phase, electrons and positrons are accelerated into two oppositely directed beams with high Lorentz factors $\gamma_{\rm b} \sim 10^7$ for real pulsars \citep{Timokhin2010}. During propagation along magnetic field lines with curvature radius $\rho_{\rm c}$, high-energy particles emit $\gamma-$rays of energy $\varepsilon$ at a rate of
\begin{equation}
    \frac{d N_{\rm ph}}{d t d \varepsilon} =  \frac{1}{\sqrt{3}\pi} \frac{e^2}{\hbar^2 c} \frac{1}{\gamma_{\rm b}^2} \int_{\frac{\varepsilon}{\varepsilon_{\rm ph}^*}}^\infty K_{5/3}(x) dx,
    \label{EmissionOfPhotons}
\end{equation}
where $K_{5/3}$ is Macdonald function, and $\varepsilon_{\rm ph}^*$ is characteristic energy of a curvature photon
\begin{equation}
    \varepsilon_{\rm ph}^* = \frac{3}{2}\hbar \frac{c}{\rho_{\rm c}} \gamma_{\rm b}^3.
    \label{PhotonEnergy}
\end{equation}
Photons are emitted tangentially to the magnetic field lines, and the angle between the photon propagation and magnetic field direction, $\psi$, increases. In our simulations, the angle increases linearly with distance from the emission point, $z$: $\sin \psi = \int {\rm d}z/\rho_{\rm c}(z)$. We do not include the displacement of photons from one field line to another during their propagation. This is justified because the transverse propagation distance, $\sim l_{\rm gap}(l_{\rm gap}/\rho_{\rm c})$, where $l_{\rm gap}$ is the longitudinal size of the gap, is significantly smaller compared to the polar cap radius, $\PC{R}$, in pulsars with $l_{\rm gap}\lesssim \PC{R}$ considered in this work.

High-energy photons propagating at a non-zero angle with respect to the strong magnetic field eventually decay into electron-positron pairs. The cross-section of the photon absorption is given by \citep{Erber1966}: 
\begin{equation}
    \frac{{\rm d}\sigma}{{\rm d}z} = 0.23 \frac{B}{B_q} \sin{\psi} \frac{\fine{\alpha}}{\lambda_c} \exp{\left(-\frac{8}{3 \chi}\right)} \Theta(\Tilde{\varepsilon}_{\rm ph} \sin{\psi}-2), 
\end{equation}
where $\chi=(B/B_q)\Tilde{\varepsilon}_{\rm ph} \sin{\psi}$, $B_q = m_e^2 c^3/e\hbar \approx 4.41 \times 10^{13}$G is the critical magnetic field, and $\Tilde{\varepsilon}_{\rm ph}=\varepsilon_{\rm ph}/m_e c^2$ corresponds to the photon energy measured in units of $m_e c^2$. The theta-function ensures that the probability of pair production is zero below the threshold, $\Tilde{\varepsilon}_{\rm ph} \sin{\psi} = 2$  \citep{Daugherty1983}. After rapid radiation of the momentum component perpendicular to the magnetic field, secondary pairs move with the following 4-velocity along the magnetic field,
\begin{equation}
    u_{||} = \frac{|\cos{\psi_a}|(\Tilde{\varepsilon}_{\rm ph}^2-4)^{1/2}}{\left(\Tilde{\varepsilon}_{\rm ph}^2 \sin^2{\psi_a}+4\cos^2{\psi_a}\right)^{1/2}} \sim \frac{1}{\sin \psi_{a}}\sim 10^2-10^3, 
    \label{SecPairs}
\end{equation}
where ${\psi_a}$ is the angle of the photon propagation direction at the point of absorption. This process generally results in large pair multiplicity, ${\cal M}=n_{\pm}/\GJ{n} \gg 1$ \citep{Timokhin2019}.

Realistic scales of particle and photon energies have so far been achieved only in one-dimensional simulations. In 2D and 3D, we adopt the following re-scaling procedure. The maximum Lorentz factor of particles accelerated in a vacuum electric field on a scale of the transverse size of the polar cap is assumed to be $\PC{\gamma} = 0.5 (\PC{R}/\UGJ{d_{\rm e}})^2$, where $\UGJ{d_{\rm e}}=c/\sqrt{4\pi |\GJ{\rho} e|/m_e}$ is the cold skin depth defined for the density of the GJ plasma. Here, we used an electric field linearly increasing with distance, appropriate for super-GJ and volume return currents (\citealt{Timokhin2013}, also see Appendix \ref{rescalingQED}). The cooling efficiency due to curvature radiation is parameterized by the radiation reaction limit, $\gamma_{\rm rad}$. It is defined as the Lorentz factor corresponding to the balance between the accelerating electric force and the radiation reaction force, $e \PC{E} = (2/3) e^2 {\gamma_{\rm rad}^4}/{\rho_{\rm c}^2}$. Finally, the energies of the curvature photons are expressed as $\Tilde{\varepsilon}_{\rm ph}=(\gamma/\gamma_{\rm emit})^3$, where $\gamma_{\rm emit}$ parameterizes the Lorentz factor of a particle emitting a curvature photon with an energy of an electron rest mass, $(3/2)\hbar ({c}/{\rho_{\rm c}}) \gamma_{\rm emit}^3 = m_e c^2$. Together with the spatial distribution of the curvature radius of field lines, and its fiducial value, $\rho_{{\rm c},0}$, and the strength of the magnetic field $B/B_{q}$, these energy scales fully determine the problem. For Vela with the rotational period $P=0.089$s and surface magnetic field strength $B_{\star} \approx 2\times 10^{12}$G, the characteristic Lorentz factors $\gamma_{\rm emit}=8\times10^5,~\gamma_{\rm rad}=1.8\times 10^8,~\PC{\gamma}=6.5\times10^9$, resulting in the hierarchy of energy scales as \mbox{$\gamma_{\rm emit} \ll \gamma_{\rm rad} \ll \PC{\gamma}$}. Our simulations downscale all these numbers (see Table~\ref{tab:sim}\footnote{The choice of the simulation parameters is made to solely downscale the gap sizes. For example, high values of the magnetic field strength, $B>B_{Q}$, lead to lower Lorentz factors of electrons/positrons that emit pair-producing photons screening the gap, $\gamma_{\rm gap}$, which allows a gap size that is smaller than our polar cap. However, we do not include any new QED-effects (e.g., photon splitting) expected for these field strengths into the simulations, as we aim to describe polar caps of rotation-powered pulsars, where these effects are negligible.}) but preserve the above hierarchy. We provide additional details of numerical implementation and QED-related rescaling relations in Appendix \ref{rescalingQED}. 

\subsection{Atmosphere}
\label{sim:atmosphere}

The surface of the NS is covered by a thin electron-ion atmosphere \cite[e.g.,][]{Haensel2006}. This atmosphere provides a reservoir of charged particles that an unscreened electric field can pull into the magnetosphere. To model the space-charge limited flow, hereafter SCLF, we place a thermal plasma just above the inner boundary of the simulation domain, where we set conducting boundary conditions for electromagnetic fields and outflow for plasma particles. The thermal plasma has a Boltzmann spatial distribution $n = n_{\rm peak} \exp{(-z/h)}$. Here, $z$ is a coordinate along the background magnetic field, $B_0$, the peak density $n_{\rm peak} \approx 10 \GJ{n}$, and $h=kT/(m_e g) \approx 10 \UGJ{d_{\rm e}}$ is a scale-height. The Boltzmann distribution is supported by adding a constant gravitational force $-m_e g$ to the particle equation of motion\footnote{We turn this force off inside the magnetosphere, since in real pulsars, the Lorentz force vastly exceeds the gravitational force.}. At each time step, the injector replenishes escaping particles within one scale height from the stellar surface to maintain the Boltzmann distribution. The temperature $T$ is selected to be low to prevent the situation when the quasi-neutral thermal outflow carries the magnetospheric current from the tail of the Boltzmann distribution \citep{Beloborodov2007}. We verify that our setup produces correct 1D SCLF solutions for both sub- and super-GJ currents in Appendix \ref{rescalingQED}. Additionally, we perform simulations adopting the RS model, where the atmospheric supply of plasma is absent. In this case, plasma is generated solely through pair production in discharge events.

\subsection{Initial plasma state}
\label{sim:initial}

Above the atmosphere, we initialize plasma with multiplicity $\cal{M}\sim {\rm few}$ to provide the magnetospheric current $j = j_{\rm mag}$ by sub-relativistic counter-streaming. This is done to avoid finding the configuration of vacuum electric fields in multi-dimensional settings and allows to start with initial conditions $\delta {\bm E} = 0$. In the 1D setup, we verified that the late time evolution of the discharge is insensitive to starting with vacuum or plasma-filled conditions. To satisfy the constraint given by Eqn. (\ref{FieldConstrain}), we add electrons, providing spatial charge density $\rho=\GJ{\rho}$. As pointed out in \cite{Timokhin2013}, gap formation in super-GJ current is sensitive to the sign of the spatial gradient of the GJ density. In particular, the gap appears close to the stellar surface if the gradient of $|\GJ{\rho}|$ is positive, as dictated by the general-relativistic correction caused by frame-dragging \citep{Beskin1990, Muslimov1992, Philippov2015}. Thus, we adopt the model of decreasing GJ charge density in the simulation box, $\GJ{\rho}=\GJ{\rho}^{0} (1 + 0.8z/L_z)$, where $L_z$ is the size of the simulation box along the magnetic field\footnote{Realistic gradient caused by the frame-dragging is significantly smaller: $|\GJ{\rho}|=(\Omega-\Omega_Z)B_0/(2\pi c)=|\GJ{\rho}^0|(1+(A/L_z) z)$, where $|\GJ{\rho}^0|=\Omega B_0(1-2C/5)/2\pi c$ and $A \approx (6C/5) (L_z/R_*)/(1-2C/5)$, where $C$ is the compactness of the neutron star. In 1D simulations performed for realistic parameters, we verified that the location of the gap is similar to more realistic values of the gradient, e.g., $A \approx 0.07$ corresponding to Vela pulsar.}.  

Simulations initialized with a uniform distribution of plasma across magnetic field lines result in a coherent discharge across the whole polar cap. For the duration of our simulations, this idealized coherence persists. In this paper, we aim to study the ``natural" state of pair discharges when discharges on all field lines do not necessarily start simultaneously, so we introduce initial inhomogeneities in plasma and let the system evolve. To desynchronize the discharges on neighboring field lines, we split the polar cap into the domains or patches. Initially, each patch is populated with a high-density plasma, $n \gg \GJ{n}$, which can be thought of as the remaining plasma from the previous discharge episode. To maintain these high-density regions, we slowly inject $e^{-}-e^{+}$ plasma into the patch to maintain $\delta {\bm E} \approx 0$ and the absence of particle acceleration. At some point, we stop the plasma injection, and the discharge restarts when most of the plasma falls into the atmosphere or escapes the box as a result of counter-streaming demanded by the magnetospheric current. We stop supplying the initial plasma at different times on neighboring patches, which leads to the desynchronization of phases of discharges. Following self-consistent evolution after all patches are cleared of the initial plasma allows us to study the phase coherence of discharges across the polar cap. The initial partition of the polar cap into patches is shown by a colored inset in Fig. \ref{fig:current}a.

\subsection{Numerical details}
\label{sim:numerical}

We perform simulations of pair production discharges using the \texttt{Tristan-v2}, multi-species radiative PIC code \citep{tristan_v2}. We present results of 2D and 3D simulations with resolution $\Delta x = \UGJ{d_{\rm e}}$, where $\Delta x$ is the cell size. In 2D, we verified that our results are unchanged for $\Delta x = \UGJ{d_{\rm e}}/2$. To ensure that the hot skin depth, $d_{\rm e}^{\rm hot} = c/\sqrt{4\pi n e^2 \left\langle 1/\gamma^3\right\rangle/m_e}$, is always resolved in the simulation domain, we limit the multiplicity of the produced electron-positron pairs at ${\cal M}_{\rm lim} = 50$. The initial magnetic field is uniform ${\bm B}_0 = B_z = {\rm const}$ in the whole computational domain. The curvature of the field lines, $\rho_{\rm c}$, is prescribed, and is only used for calculating emission of photons and pair production. The distribution of the field-aligned component of the magnetospheric current is constant along magnetic field lines, as is the case in stationary FFE solutions. Our system size is $\PC{R}/\UGJ{d_{\rm e}}=8\times10^3 \div 1.2\times10^4$ for 2D, resulting in the Lorentz factors corresponding to the full vacuum potential drop across the polar cap, $\PC{\gamma} = 0.5 (\PC{R}/d_{\rm e})^2 \approx 5\times10^7$. In 3D, the size of the polar cap is $\PC{R} = 10^3 \UGJ{d_{\rm e}}$, and correspondingly $\PC{\gamma} \sim 10^6$. The simulation domain size is $2 \PC{R}$ along the stellar surface and $2 R_{\rm PC}$ along the field lines. We study both the SCLF and RS models with varying parameters, corresponding to small, $l_{\rm gap} \ll \PC{R}$, and large, $l_{\rm gap} \lesssim \PC{R}$, gap sizes in 2D. Additionally, we perform a simulation in the SCLF model of a marginally small gap, corresponding to the current distribution in the 60$^\circ$-inclined rotator. The summary of the parameters of different simulations is given in Table \ref{tab:sim}. We employ $8$ filter passes on the deposited currents to reduce the particle noise. The fields are evolved using modified stencils \citep{Blinne2018}, which significantly improve the isotropy of the numerical dispersion relation of the electromagnetic waves. This method also leads to a substantial reduction of the electromagnetic noise in our multi-dimensional simulations compared to the evolution employing the standard Yee solver.

\section{Results}
\label{results}

\begin{figure*}[t]
\includegraphics[width=\textwidth]{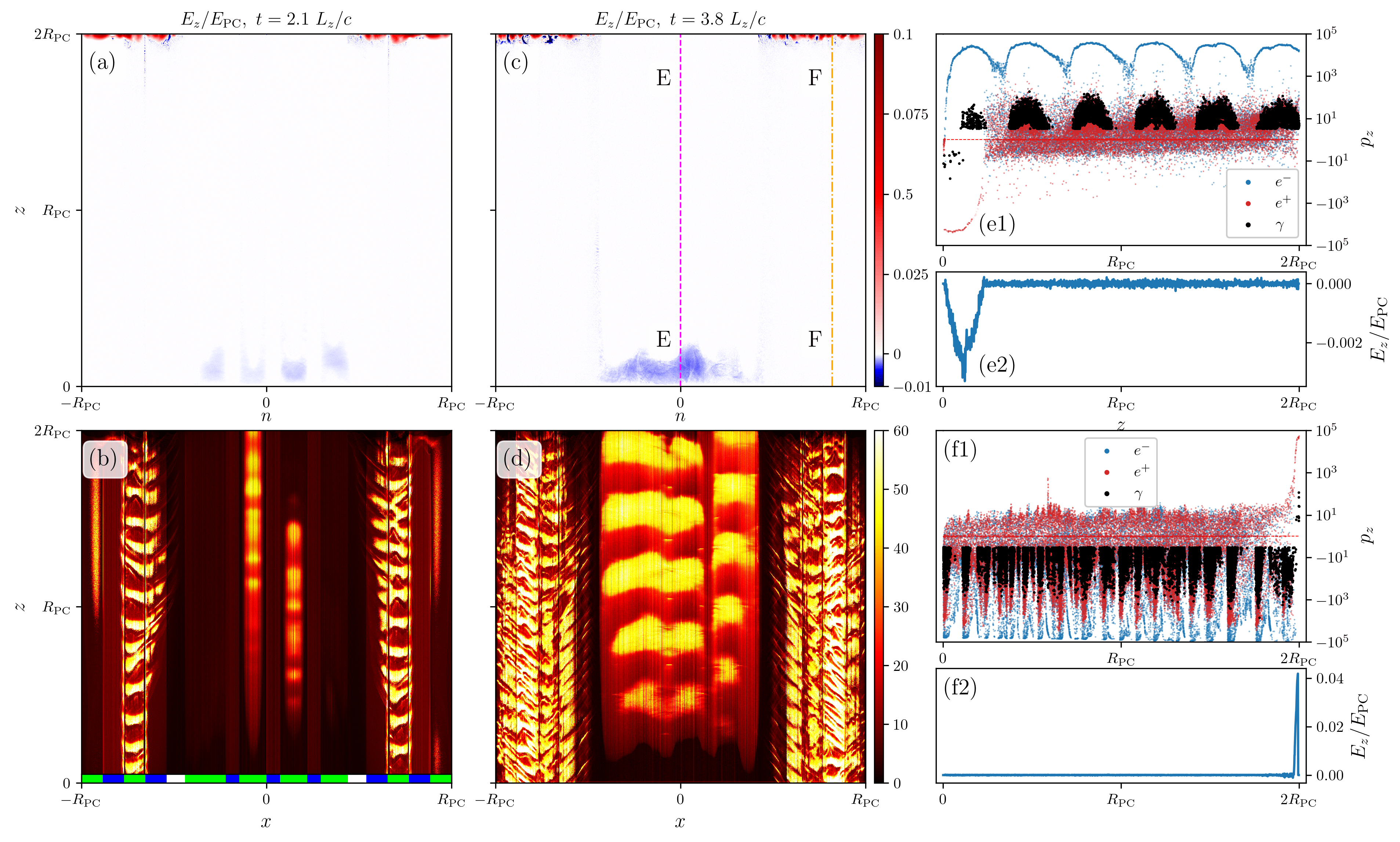}
\caption{Evolution of the discharge in the 2D polar cap with a small gap, $l_{\rm gap} \ll \PC{R}$, and free particle escape from the stellar surface (SCLF). The first column depicts (a) the electric field and (b) plasma density during the initial phase of the evolution, where patches occupy a large area of the polar cap. At this time, the discharge along the super-GJ field lines behaves in an almost steady-state manner. The second column illustrates a later evolutionary stage when all patches are cleared of the initial plasma, leading to a cyclic discharge. The asymmetry of clouds of produced plasma suggests the decorrelation of discharges on neighboring field lines. Lines ``E'' and ``F'' indicate the positions of 1D cuts of the $z-p_z$ phase space and electric field for the region of the super-GJ current  (e1,e2) and (f1,f2) - for return current. In the phase space plots, blue, red, and black dots represent the position and momentum of electrons, positrons, and pair-producing photons, correspondingly.}
\label{fig:SmallSCLF}
\end{figure*}

1D simulations in the SCLF regime found qualitatively different regimes of the discharge operation, dependent on the value of the magnetospheric current. \citep{Timokhin2013}. 
\begin{enumerate}[label=(\roman*)]
    \item Super-GJ current, $j_{\rm mag}/\GJ{j} > 1$. The gap opens close to the stellar surface. A beam of high-energy electrons is accelerated outwards by the electric field in the gap and produces photons and secondary pairs that escape into the magnetosphere. Positrons in the gap accelerate towards the star and produce pairs bombarding the surface.
    \item Sub-GJ current, $0 < j_{\rm mag}/\GJ{j} < 1$. Here, a sub-relativistic beam of electrons of density $\approx n_{\rm GJ}$ is extracted from the atmosphere, supplying enough charge density and current to screen the electric field. No pair production happens in this zone. 
    \item Return current, $j_{\rm mag}/\GJ{j} < 0$. The gap opens at high altitudes, close to the upper boundary of the simulation domain\footnote{It is likely that this gap initially opens at the null-surface, where $\GJ{\rho} =0$, see evidence in global simulations in, e.g., \cite{Bransgrove2023}.}. Since the gap is located at a large distance from the star, the most efficient pair production is driven by the particles, emitting curvature photons towards the region of stronger magnetic field, i.e., propagating to the stellar surface. At the same time, the discharge is ignited by photons emitted by positrons moving outward (see \cite{Timokhin2010} and Section \ref{sec:SCLFsmall}).  
\end{enumerate}

In the RS model, \cite{Timokhin2010} demonstrated that the magnetic field lines with the $j_{\rm mag}/\GJ{j} > 1$ and $j_{\rm mag}/\GJ{j} < 0$ operate overall similarly to the SCLF case. On the other hand, field lines with sub-GJ current $0 < j_{\rm mag}/\GJ{j} < 1$ become capable of particle acceleration and subsequent pair production. Our multi-dimensional simulations support these basic qualitative conclusions but also reveal novel dynamics across magnetic field lines.

\subsection{Space-Charge-Limited Flow}
\label{sub:sclf}

In this Section, we present the results of simulations of a discharge in the 2D polar cap with the supply of plasma from the stellar atmosphere (see simulations \texttt{SCLFsmall}, \texttt{SCLFsmallCurv}, \texttt{SCLFsmallTail}, \texttt{SCLFlarge} in Table~\ref{tab:sim}). In all of these simulations, we start with three patches loaded with low-density plasma, allowing quick opening of the gap on these field lines. Two of the patches are located in the zone of return current, and one is in the zone of the super-GJ current.  

\subsubsection{Small Gap, Constant Curvature of Field Lines} 
\label{sec:SCLFsmall}

In polar caps of real pulsars, the magnetic field can deviate from dipolar configuration due to the presence of local multipolar components, resulting in magnetic field configurations that do not necessarily possess axial symmetry around the center of the polar cap. Our configurations with constant curvature radius represent such polar caps.
The results of the simulation \texttt{SCLFsmall} with a small gap, $l_{\rm gap}\ll \PC{R}$\footnote{Sizes of gaps are quoted for the zone with super-GJ current, in the state when all patches are cleared of the initial plasma. Parameters of the simulation in Table~\ref{tab:sim} are calibrated using 1D simulations for corresponding values of $j_{\rm mag}$.}, and constant field line curvature, $\rho_{\rm c}= \rho_{{\rm c},0}$, are shown in Fig. \ref{fig:SmallSCLF}. In the super-GJ zone, the gap forms close to the stellar surface. Initially, electric potential along the gap is limited by the transverse size of the patch, resulting in slow acceleration and low energies of electrons extracted from the atmosphere. The mean free path of photons emitted by these electrons is large, and pair production happens beyond the gap, i.e., in a region with $\delta {\bm E} \approx 0$. Since produced pairs have a non-zero momentum directed outward, none of them can return into the gap zone to screen the electric field. At the same time, the region beyond the gap is filled with high-density secondary plasma, preventing the gap from further expansion. Under these conditions, the gap becomes quasi-stationary: its size does not change in time. We find that the gap operates in a quasi-stationary regime when the size of the gap along the field lines is large, $l_{\rm gap} \gg w$, where $w$ is the transverse width of the patch. Quasi-stationary gaps are seen in the active patches in the zone with super-GJ current at early times, as shown in the left column of Fig.\ref{fig:SmallSCLF}.

When a new patch between the two already active patches with quasi-stationary gaps is cleared of plasma, the unscreened electric field in the plasma-starved zone communicates across the field lines. This leads to a significant increase in the accelerating potential, proportional to the increase of the transverse size of the gap. Electrons, accelerated to higher energies, emit more energetic photons that are absorbed within the gap, leading to the cyclic screening of the electric field. The example of the merging of a few quasi-stationary gaps can be seen in the central column of Fig.\ref{fig:SmallSCLF} (subplots c-d). These snapshots correspond to the later time when all patches are cleared of initial plasma, and the whole polar cap is available for discharge. 

In a cyclic discharge, the unscreened electric field on a particular field line exists during $\tau_{\rm active} \approx l_{\rm gap}/c$. Thus, the transverse size of the region affected by the unscreened electric field on the field line cannot exceed $c\tau_{\rm active} = l_{\rm gap}$. Hence, if two field lines are separated by a distance exceeding $2 l_{\rm gap}$, their electric fields can not communicate. This condition sets the limitation to the transverse coherence scale of a discharge, $l_{\perp}\sim l_{\rm gap}$. The transverse variation of shapes of the front of the unscreened electric field and clouds of secondary plasma seen in Fig.\ref{fig:SmallSCLF}d present a hint of the desynchronization of discharges at different field lines. We note that the size of the cyclic gap on field lines carrying the super-GJ current is larger compared to the one expected for a vacuum electric field. Here, the electric field extracts electrons from the stellar atmosphere and positrons from the plasma cloud produced in the previous discharge episode into the gap zone. The presence of both electrons and positrons results in a smaller deviation of the plasma charge density from the GJ value, leading to smaller electric field in the gap, $\sim 0.1 |E_{0}|$, where $E_{0} = 4\pi \GJ{\rho}l_{\rm gap}(|j_{\rm mag}/\GJ{j}| - 1)$ corresponds to the case of only electrons carrying the current (see the 1D slice of the electric field in subplot (e2) of Fig.\ref{fig:SmallSCLF}). 

On field lines carrying the return current, a strong electric field appears at the upper boundary of the simulation box, where a lack of electrons develops. The current in the gap zone is carried by positrons moving into the magnetosphere. This leads to strong accelerating electric fields reaching \mbox{$E_{z}\sim E_{0} = 4\pi |\GJ{\rho}|l_{\rm gap}(|j_{\rm mag}/\GJ{j}| + 1)$}. The electric field in the return current zone\footnote{Note that the color bar in the electric field panels (subplots a and b) has different lower and upper bounds.} is shown in the subplot (f2)  of Fig.~\ref{fig:SmallSCLF}. Photons emitted by the high-energy positron beam decay into pairs moving into the magnetosphere. The unscreened electric field reverses some of the electrons towards the star and accelerates them to high energies, igniting the discharge moving inwards (see a 1D cut of the phase space $z-p_z$ in subplot (f1) of Fig.~\ref{fig:SmallSCLF}). At the same time, particles moving into the magnetosphere leave the zone of a strong enough magnetic field to allow single photon conversion, and pair production driven by these particles ceases. Because of the much stronger electric fields, the gap size in the return current zone is smaller compared to the gap in the super-GJ zone. Since here $l_{\rm gap}\ll w$, the discharge within the first open patch is immediately in an intermittent regime (subplot (b) of Fig.\ref{fig:SmallSCLF}), for chosen parameters of the gap and widths of initial patches\footnote{At the same time, the discharge in the outermost isolated patches initially occurs in a quasi-stationary regime due to the low available potential on field lines with small return currents. It remains stationary until the patch next to it becomes available for the discharge. After a neighboring patch opens, the electric fields of the two patches synchronize, and discharge continues intermittently.}. When all patches are cleared of initial plasma, the discharge proceeds intermittently on all field lines. The complex shape of the clouds of secondary plasma additionally points to the transverse desynchronization of the gaps at the scale of $l_{\perp} \sim l_{\rm gap}$. 

\begin{figure*}
\includegraphics[width=\textwidth]{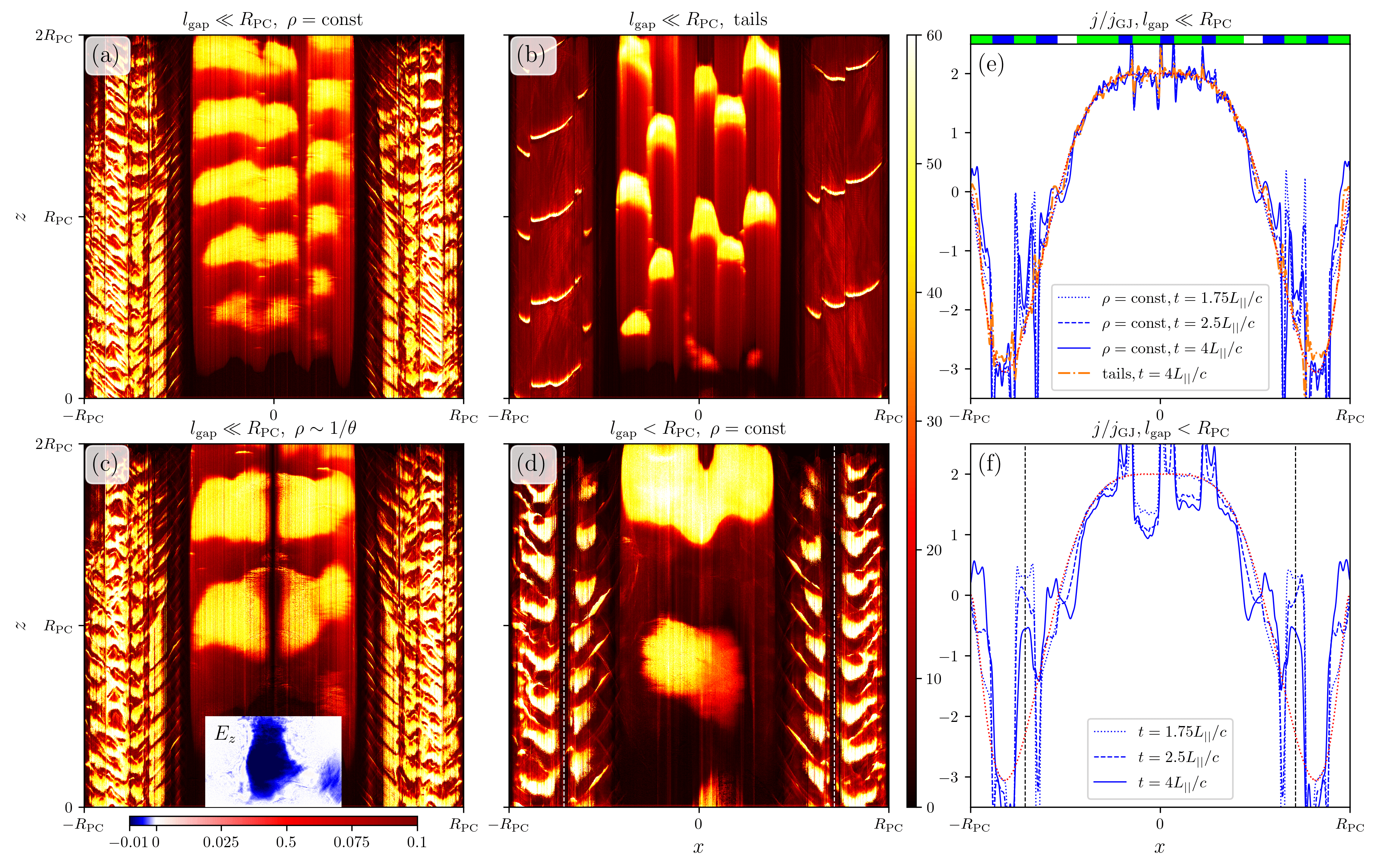}
\caption{Comparision of plasma density distributions for various simulations in the SCLF regime. (a) Small gap $l_{\rm gap} \ll \PC{R}$ with constant curvature (same as used in Fig. \ref{fig:SmallSCLF}, simulation \texttt{SCLFsmall} from Table \ref{tab:sim}). Panel (b) shows significant decorrelation of discharges on different field lines for a small gap and artificially reduced repetition rate (simulation \texttt{SCLFsmallTail}). Panel (c) presents the simulation \texttt{SCLFsmallCurv} with the same parameters but with dipolar curvature, $\rho_{\rm c} \sim 1/\theta$. The inset displays the field-aligned component of the electric field, highlighting the inclined fronts of field screening. Snapshot (d) depicts the case of a large gap, $l_{\rm gap} \lesssim \PC{R}$ (\texttt{SCLFlarge}). The right column shows plasma currents averaged along the field lines at different times for (e) small and (f) large gaps, indicating that smaller gaps more effectively sustain the twist of the magnetospheric field lines. The dotted red line in subplots (e,f) shows the distribution of the magnetospheric current density.}
\label{fig:FullSCLF}
\end{figure*}

\subsubsection{Small Gap, Strong Desynchronization}
\label{sec:SCLFtails}

Both regular \citep[e.g.,][]{Manzali2007, Rigoselli2022} and millisecond \citep{Miller2019, Miller2021} pulsars show thermal X-ray emission from their surfaces. The source of heating behind this radiation is thought to be the bombardment of the stellar surface by the ultra-relativistic beam of particles accelerated in a gap. These backward propagating beams can be seen in 1D cuts of phase spaces $z-p_z$, shown in subplots (e1)-(f1) of Fig.~\ref{fig:SmallSCLF}. During the active phase of the cyclic discharge, the densities of counter-streaming beams are approximately equal to each other, and the X-ray luminosity can be estimated as $L_X = 0.5 L_0 (\gamma_{\rm gap}/\PC{\gamma})$, where $L_0$ is a spin-down luminosity and $\gamma_{\rm gap}$ is the maximum Lorentz factor of particles in the gap. It is long known that this estimate gives significant overheating of the stellar surface compared to the observations \citep[e.g.,][]{Beskin2018}. In stationary gap models, derived for nearly GJ currents, the amount of particles bombarding the star is much less compared to the outflow, and the surface heating is reduced (see, e.g., \citep{Arons1981, Harding2001} and discussion in Section~\ref{sec:SCLFsmall}.) For the cyclic discharge, the surface heating is attenuated by a repetition rate, $f=\tau_{\rm active}/\Delta\tau_{\rm{rep}} \ll 1$, where $\tau_{\rm active}$ is again the lifetime of the unscreened electric field, and $\Delta\tau_{\rm{rep}}$ is the time between successive discharges \citep{Timokhin2015}.

We expect some long-lived oscillations of an electric field to remain inside the secondary plasma cloud after the main screening event in a gap \citep{Tolman2022}. These oscillations can reverse a fraction of injected secondary particles and deliver them to the gap zone. Suppose the process of particle reversal persists for a long time, $\gg \tau_{\rm active}$. In that case, the electric field behind the cloud of secondary plasma remains screened, and the repetition rate $f$ is small. 

In our multi-dimensional simulations, we observe $\Delta\tau_{\rm{rep}} \approx \rm{few} \times \tau_{\rm active}$, which is not sufficient to explain the X-ray observations. We believe this is because of the low allowed plasma multiplicity in the clouds and, hence, low-density backflow -- a conjecture that we will address in the forthcoming work. In order to decrease the repetition rate, we inject additional extended tails behind the escaping clouds of secondary plasma. We trace high-density clouds of curvature photons and inject additional pairs in cells where the photon cloud passed less than $0.3 L_z/c$ ago. The parameters of the simulation are chosen such that the discharges are in the cyclic regime within each initial patch, which corresponds to an even smaller gap size compared to \texttt{SCLFsmall} and stronger desynchronization. As shown in Fig. \ref{fig:FullSCLF}b, simulation \texttt{SCLFsmallTail} based on this prescription indeed results in long tails and small repetition rates. While the size and the lifetime of initial patches are identical to ones in simulation \texttt{SCLFsmall}, the structure of discharges is substantially different (the density snapshot for \texttt{SCLFsmall} is reproduced in Fig.~\ref{fig:FullSCLF}a for convenience). Discharges happen completely independently within each transverse zone $l_{\perp}\sim l_{\rm gap}$ with no synchronization between each other. We coin the resulting structure as a ``lava lamp'' discharge. This result follows from our previous statement that synchronization of neighboring discharges proceeds during overlapping plasma-starved episodes. The probability for two initially desynchronized gaps to appear next to each other in neighboring patches is $\approx f$. Hence, for low repetition rates, $f\ll 1$, the probability of spontaneous synchronization is very low. At the same time, the discharge remains coherent within one patch of size $\sim l_{\rm gap}$. In our simulations, this coherence is preserved by the weak variation of the repetition rate across magnetic field lines.

\subsubsection{Small Gap, Quasi-Dipolar Field}
\label{sec:SCLFdipole}

For the case of a commonly assumed star-centered dipole, the field line curvature diverges, $\rho_{\rm c} \approx 4\sqrt{R_\star \LC{R}}/3\theta$, where $\theta$ is the polar angle, making pair production impossible near the magnetic axis. To study discharges in a dipolar geometry, in simulation \texttt{SCLFsmallCurv} we impose the curvature of the field lines in the form $\rho_{\rm c} = \rho_{{\rm c},0} ({\PC{R}}/{x})$, where values of $\rho_{{\rm c},0}$ and $\PC{R}$ are given in Table~\ref{tab:sim}, and $x$ is the distance from the magnetic pole (center of the polar cap).

 In Fig.~\ref{fig:FullSCLF}c, we show a snapshot of the plasma density in the simulation. We find that close to the edge of the polar cap, in the return current zone, the discharge behavior is very similar to the simulation \texttt{SCLFsmall}. The field lines carrying the super-GJ current, however, have a substantially larger radius of curvature than the outer field lines. Comparison of snapshots with constant and dipolar curvatures, Fig.~\ref{fig:FullSCLF}a,c, demonstrates that the gap in the super-GJ current zone is overall larger in the simulation with dipolar curvature than in the simulation with the constant curvature. This behavior can be easily understood considering the analytical scaling for the gap size \citep{Timokhin2015}:
\begin{equation}
    l_{\rm gap} \propto \xi^{-3/7} \rho_{\rm c}^{2/7} P^{3/7} B^{-4/7},
    \label{gapSize}
\end{equation}
where $\xi\approx j_{\rm mag}/\GJ{j}$ characterizes the strength of the electric field, $P$ is the pulsar's rotational period, and $B$ is the magnitude of the surface magnetic field. The variation of the gap size across the field lines in the super-GJ zone is determined by two factors: the strength of the accelerating electric field and the radius of curvature of the field lines. Since both $\xi\approx j_{\rm mag}/\GJ{j}$ and the radius of curvature increase towards the pole, these two effects nearly cancel each other. Consequently, we find no substantial variation in the gap size everywhere except in the region very close to the magnetic axis where the field curvature diverges. This result coincides with early theoretical conclusions \citep[e.g.,][]{Arons1979}. Here, the gap size becomes infinite because the energy of curvature photons is very low (see Eq.~\ref{EmissionOfPhotons}), and a hole in the plasma density forms.

The desynchronization of the discharges on neighboring field patches in the return current zone is practically identical to the simulation \texttt{SCLFsmall} (see Section \ref{sec:SCLFsmall}). Synchronization of the discharges in the super GJ-current zone is more active due to the overall larger size of the gap (also see Section~\ref{sec:SCLFlarge}). Here, synchronization of two patches on opposite sides of the magnetic axis occurs during the plasma-starved episodes because the unscreened electric field always exists in the central plasma density hole. Since the gap size varies substantially in a narrow region close to the magnetic axis, the front of the electric field screening is highly inclined with respect to the magnetic field, as shown in the inset of Fig. \ref{fig:FullSCLF}c. As pointed out by \cite{Philippov2020, Cruz201}, these inclined electric field screening fronts are plausible sources of superluminal ordinary modes, which can transform into the observed radio emission as they escape from the magnetosphere.

\subsubsection{Large Gap}
\label{sec:SCLFlarge}

\begin{figure*}
\includegraphics[width=\textwidth]{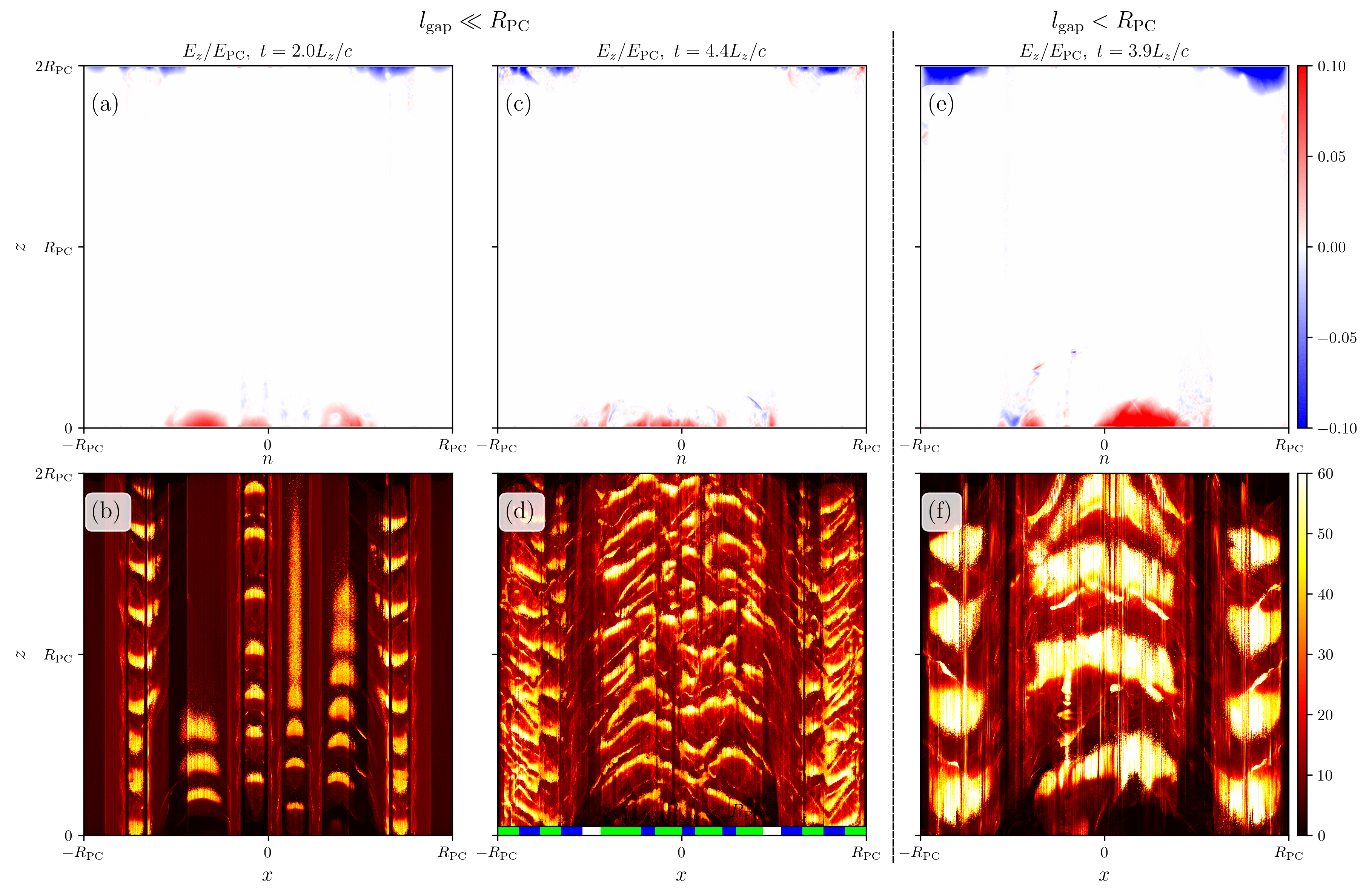}
\caption{Evolution of the discharge in the 2D RS polar cap with a small, $l_{\rm gap}\ll \PC{R}$, \texttt{RSsmall}, and large, $l_{\rm gap}\lesssim \PC{R}$, \texttt{RSlarge}, gaps. The first column shows (a) electric field and (b) plasma density distribution during the initial phase of the evolution in \texttt{RSsmall}, when initial patches still occupy a large area of a polar cap. The second column, panels (c)-(d), illustrates a later evolutionary stage when all patches are cleared of the initial plasma. The right column shows (e) the electric field and (f) density snapshots for a large gap, \texttt{RSlarge}, at a later time when the initial patches are cleared of plasma.}
\label{fig:FullRS}
\end{figure*}

In this Section, we study a discharge with a gap size comparable to the size of the polar cap $l_{\rm gap} \sim \PC{R}$, corresponding to less energetic pulsars. In Fig.~\ref{fig:FullSCLF}d, we show a snapshot of the plasma density distribution in the simulation \texttt{SCLFlarge}. At early times, a few isolated narrow patches are available for discharge in the region with the super-GJ current. In each of the patches, a quasi-stationary gap forms as described in Section \ref{sec:SCLFsmall}. After all patches are cleared of the initial plasma, the discharges on neighboring field lines synchronize across most of the super-GJ zone of size $l_{\perp}\sim l_{\rm gap}$. The resulting coherent discharge proceeds intermittently.

Similar conclusions apply to the discharge in the return current region, where we observe a pair of coherent discharges in each zone, see Fig.~\ref{fig:FullSCLF}d. They are separated by a plasma density hole (white dashed lines in Fig.~\ref{fig:FullSCLF}d mark its center), which has a peculiar origin. Here, if the patch width is too small to establish a high enough electric potential to trigger pair production, the whole field line can become clear of pairs because of the counter-streaming demanded by the magnetospheric current. Simultaneously, an electric field extracts massive ions from the stellar atmosphere, which are not able to emit high-energy photons. In our simplified setup that does not take the transverse motion of photons into account, zones that once become clear of pairs can no longer ignite a discharge. These field lines possess an unscreened electric field that allows transverse communication of neighboring patches.

\subsection{Ruderman-Sutherland Gap}
\label{sub:RS}

The SCLF model \citep{Arons1979} corresponding to the free escape of particles from the stellar atmosphere seems more physically motivated; however, the RS model with no plasma supply from the star is still widely used to explain various observations \citep[e.g.,][]{Wen2020, Primak2022, Janagal2023}. 
In this Section, we consider discharges in the model with no plasma supply from the stellar surface (RS). The profile of the magnetospheric current and fragmentation of the polar cap into patches is identical to the simulations in the SCLF model, as described in Section~\ref{sec:SCLFsmall}. The RS model is typically applied to a pulsar with counter-aligned magnetic moment and angular velocity vectors, the ``anti-pulsar''. In this geometry, the GJ density is positive. In comparison with SCLF simulations, we change the sign of both the GJ density (see Section \ref{sim:initial}) and the magnetospheric current. 

In Fig. \ref{fig:FullRS}, we show snapshots of the accelerating electric field and plasma density for the simulations with small (simulation \texttt{RSsmall}) and large (\texttt{RSlarge}) gaps. The evolution of discharges in the RS model is very similar to that of the SCLF model: the discharges operate in an intermittent regime, and the directions of discharges are identical to those in the SCLF polar cap. Namely, the gap opens at the stellar surface, and clouds of secondary plasma move toward the magnetosphere in zones of the positive current. In the regions of return current, a gap opens at high altitudes and launches plasma moving toward the star. It is important to note that the area with sub-GJ magnetospheric current also possesses an accelerating electric field and pair-producing discharge, which is in contrast with the SCLF model. The RS and SCLF models also differ in the zones of the super-GJ current, where the RS model has a gap with large parallel electric field $\approx E_{\rm 0}\approx 4\pi \GJ{\rho} l_{\rm gap}$. In the RS model, the current is carried by particles of one sign extracted from the cloud created during the previous discharge. In the SCLF model, the electric field is much smaller because particles of both signs are present in the gap. 

In the case of a small gap, $l_{\rm gap}\ll \PC{R}$, Fig.~\ref{fig:FullRS}a-d, the discharge operates in the intermittent regime already when the first patches are cleared of the initial plasma (Fig.~\ref{fig:FullRS}a,b), in the zones of both super-GJ and return currents. At a later time, when all of the patches are cleared, the discharges communicate through the vacuum episodes similar to the discharges in the SCLF model. The complex shape of the clouds of secondary plasma demonstrates that the discharges are not fully synchronized across magnetic field lines. At the same time, a relatively large repetition rate, $f$, leads to smaller desynchronization of the discharges (Fig.~\ref{fig:FullRS} c,d). We expect a similar ``lava lamp'' pattern of de-synchronized discharges in a situation of a realistically low repetition rate, $f \ll 1$, as described in Section~\ref{sec:SCLFtails}. In the polar cap with a large gap, \texttt{RSlarge}, $l_{\rm gap} \lesssim \PC{R}$, see Fig.~\ref{fig:FullRS}e-f, a quiet stage at the beginning of simulations with narrow, isolated patches is later replaced by a well-developed cascade with nearly synchronized discharges when all patches are cleared of the initial plasma. This dynamics is broadly similar to the evolution in the SCLF simulation with a large gap, \texttt{SCLFlarge}.

\subsection{3D Gap, Space-Charge-Limited Flow}
\label{sec:3D}

\begin{figure*}
\includegraphics[width=\textwidth]{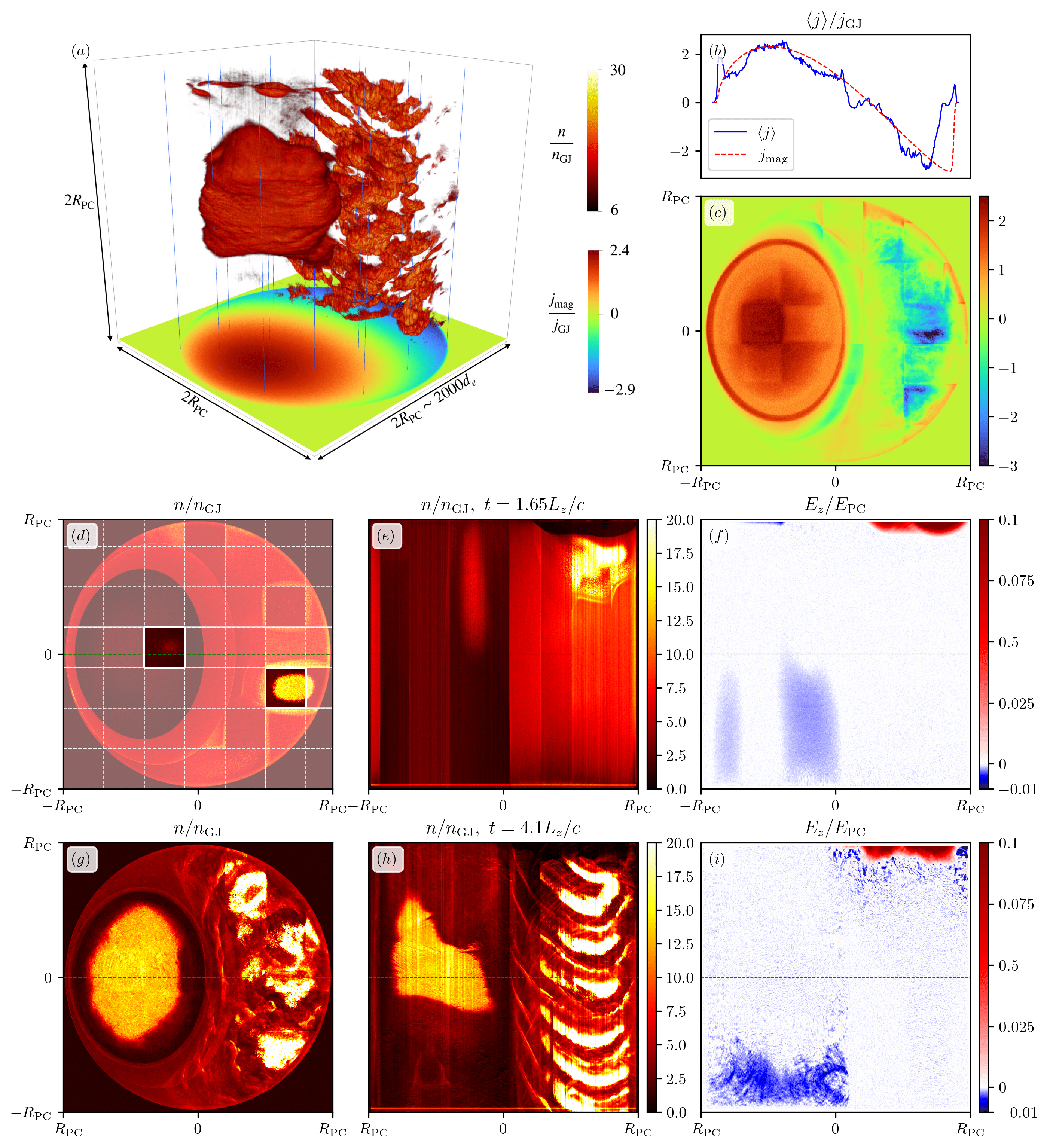}
\caption{3D SCLF discharge in a polar cap of a 60$^\circ$-inclined rotator with a marginally small gap, $l_{\rm gap} < \PC{R}$. Subplot (a) shows a volume rendering of the density of a secondary plasma. The plane at the bottom shows the position of the stellar surface, and the colormap shows the distribution of the magnetospheric current. The blue lines perpendicular to the stellar surface represent the magnetic field lines. Subplots (b) and (c) show the time-averaged distribution of the plasma current. The black dashed line in the plot (c) shows the position of the 1D slice in (b). Two lower rows show 2D slices of the plasma density (d,e,g,h) and electric field (f, i) through the center of the domain in two different projections. The dashed lines in each of the slices show the position of a slice in a perpendicular plane, e.g., the dashed line in (d) shows the position of slices (e) and (f). Snapshots (d-f) in the middle row show the initial phase of the simulation, when a lot of patches are still filled with initial plasma, and the lower row (g-i) shows a time when all patches are cleared of the initial plasma. White dashed lines in subplot (d) show the boundaries of the initial patches, white shadowed regions show initially inactive patches and 2 not-shadowed regions represent initially active patches.}
\label{fig:3dPolarCap}
\end{figure*}

In this Section, we study a 3D discharge corresponding to the current distribution in the $60^\circ$-inclined rotator. Here, almost half of the polar cap is occupied by the field lines with super-GJ current, and the other half is occupied by the field lines with negative return current (see Fig.\ref{fig:current}b). The choice of the simulation parameters (simulation \texttt{SCLF3D} in Table \ref{tab:sim}) provides a small gap, $l_{\rm gap} \ll \PC{R}$, in the region of the return magnetospheric current, and marginally small gap, $l_{\rm gap} < \PC{R}$, in the region of super-GJ current, due to different electric potentials developing in these two zones (see Sec.~\ref{sec:SCLFsmall})\footnote{Further increase of the separation of scales, $\PC{R}/l_{\rm gap}$, is challenging because of the computational cost of the simulations.}. To study the desynchronization of the discharge, we split the polar cap into $6\times6$ square patches, with two of them being 
initially loaded with low-density plasma, which allows the formation of a gap at the beginning of the simulation. One of two ``open'' patches is located in the zone of super-GJ current, and another one - is in the region of return current (the positions of these two patches are shown in Fig. \ref{fig:3dPolarCap}d). In other patches, up until certain time moments, we inject additional plasma to prevent charge starvation and to keep the accelerating electric field screened, as discussed in Section \ref{sim:initial}.

Overall, the evolution of the discharge is similar to the 2D SCLF simulations discussed in Section~\ref{sec:SCLFsmall}. The volume rendering of the plasma density for a well-developed discharge is shown in Fig. \ref{fig:3dPolarCap}a. In the region of the super-GJ current, the transverse size of the patch limits the electric potential, leading to the formation of a large quasi-stationary gap near the stellar surface. The examples of such two isolated gaps can be seen in the early time slice of the electric field, $E_z$, Fig.~\ref{fig:3dPolarCap}f. The corresponding slice of the plasma density, Fig.~\ref{fig:3dPolarCap}e, shows the production of secondary plasma beyond the gap. At later times, other patches are cleared of the initial plasma and become available for discharge. Similar to 2D large gap simulations in Sec.~\ref{sec:SCLFlarge}, neighboring discharges synchronize on transverse scales $l_\perp \sim l_{\rm gap}$, leading to a nearly coherent discharge across the super-GJ zone, as clearly visible in a slice of the electric field in Fig.~\ref{fig:3dPolarCap}. We start with only two relatively small patches where discharges are possible and observe that the discharge zone spreads across the polar cap when we stop suppressing pair formation. In a well-developed cascade, discharges produce large clouds of secondary plasma, which spread across all field lines, as shown in Fig.~\ref{fig:3dPolarCap}g,h. 

The formation of the gap in the region of the return current starts at high altitudes. Similarly to a 2D discharge, a lack of electrons develops there, leading to a strong accelerating electric field, $E_z \sim E_0 \sim 4 \pi |\GJ{\rho}| l_{\rm gap}$. It leads to a smaller gap compared to the super-GJ field lines, and hence, a cyclic discharge can start within each initial patch. When all patches are cleared of the initial plasma, significant desynchronization of the discharges across magnetic field lines remains, as seen in the transverse slice of the plasma density, Fig.~\ref{fig:3dPolarCap}g. Here, a few isolated structures are seen, similar to the 3D volume rendering in Fig.~\ref{fig:3dPolarCap}a. 

\subsection{Evolution of the Twist of Magnetic Field Lines}
\label{sec:curEvol}

As was discussed in Section~\ref{GapDynamics}, the persistent twist of the magnetic field can be maintained in the plasma-filled, i.e., FFE, state only. This condition is naturally broken in the gap zone. The lack of plasma can have two consequences: the growth of the accelerating electric field and the untwisting of the field lines (increasing $\partial{\delta\bm{E}}/\partial{t}$ vs. decreasing  $\nabla{}\times\delta\bm{B}$ terms in Eq.~\ref{FieldEvolution1}). If magnetic field lines become untwisted in a significant fraction of the polar cap, the outflow of plasma along them would be similar to the free outflow into the vacuum, as it was implicitly assumed to be the case in original polar cap models of \cite{Arons1979} and \citet{Ruderman1975}: these models disregarded the presence of the magnetosphere in their estimates of the accelerating electric field. Such an outflow would be ``charge-density-driven''. The polar cap and the magnetosphere are parts of the same electric circuit. In this circuit, the magnetosphere, which has a much larger inductance, determines the current density along magnetic field lines. In a real pulsar, if the magnetic field lines become untwisted in the large portion of the polar cap, an ``untwisting wave'' will propagate into the magnetosphere, where it will be eventually stopped, and a new wave will be sent toward the polar cap, restoring the twist.  In our setup, we model the presence of the magnetosphere by imposing the current $j_{{\rm mag}}$ in the hope that the system has an inductance high enough to maintain the prescribed current density. A situation where we end with almost a complete untwisting of magnetic field lines would mean that our setup does not model the system polar cap + magnetosphere correctly, which basically means that our simulation domain is not large enough. Therefore, checking for untwisting of magnetic field lines is an important consistency test of our model.

To study the evolution of the twist of magnetic field lines in our simulations, we compare the plasma current, $j$, with the prescribed magnetospheric current, $j_{\rm mag}$. 
In the right column of Fig.~\ref{fig:FullSCLF}, we compare the plasma and magnetospheric currents for simulations with different ratios of the gap height to the polar cap size. For small gaps, a significant fraction of any field line is loaded with high-multiplicity plasma, leading to the deviations of $j$ from $j_{\rm mag}$ to be localized in a small volume. The magnetospheric twist is easily sustained by the pair production discharge in the polar cap zone, as shown in subplot \ref{fig:FullSCLF}e. The twist is preserved particularly well in the ``lava-lamp'' discharge (simulation \texttt{SCLFsmallTails}) since the gaps, where FFE conditions are violated, are small and most of the domain is filled with dense plasma most of the time. In contrast, the simulation \texttt{SCLFsmall} starts with initially long, $l_{\rm gap} \gtrsim l_{\perp}$, ``charge-density'' driven, quasi-stationary gaps that are formed within narrow patches, as described in Section~\ref{sec:SCLFsmall}. These large gaps lead to a noticeable untwist of the field lines. This effect is particularly strong at the edges of open patches, where the electric potential is smaller due to proximity to the neighboring plasma patch, and, thus, particle acceleration and pair production are less efficient. However, when neighboring patches are cleared of the initial plasma, the electrostatic potential quickly increases, leading to a smaller gap size along the field lines. At this time, the gap becomes effectively shorter along the transverse dimension, $l_{\rm gap}\lesssim l_{\perp}$. The domain fills with dense plasma before the complete untwisting of field lines, the initial magnetospheric twist is partially restored, and discharges enter into the ``current-driven'' phase. This can be seen in the subplot \ref{fig:FullSCLF}e as a decrease in the amplitude of the difference between the plasma current and the magnetospheric one, which is particularly noticeable at the boundaries of initial patches. The evolution of the twist in the polar cap with a large gap, simulation \texttt{SCLFlarge}, is overall similar to the evolution of the small gap case. The main difference is in a more significant deviation of the plasma current from the magnetospheric current, as shown in subplot \ref{fig:FullSCLF}f, because of the larger region sustaining parallel electric fields. Patches with a low plasma current (centered around the dashed lines in Fig.~\ref{fig:FullSCLF}d,f) end up as non-sustaining the discharge, as described in Section~\ref{sec:SCLFlarge}. The field lines within these patches partially restore their twist due to the supply of plasma from the atmosphere. 

Similar conclusions hold in the case of 3D discharges, as is shown in Figs \ref{fig:3dPolarCap}b, c, where it is clear that the distribution of the plasma current overall follows the magnetospheric current. The deviations are localized to the boundaries of the patches, as clearly visible in Fig.~\ref{fig:3dPolarCap}c. Following the results of the twist evolution over a longer time in 2D simulations, we assume that the twist will be gradually restored, and the clearly visible patch boundaries will disappear over longer times (see Section~\ref{sec:SCLFsmall}).

Overall our numerical setup is capable of dealing with the untwisting of magnetic field lines. It never comes to complete untwisting on field lines, and the average plasma current is reasonably close to the prescribed magnetospheric current, $j_{\rm{mag}}$.
The size of the simulation domain is adequately large to simulate the presence of a large magnetosphere above the cascade zone. On the other hand, modeling of discharges with gap heights larger than the size of the polar cap, $l_{\rm{gap}}\gtrsim R_{\rm{pc}}$, would require a substantially larger computational domain to prevent the complete untwisting of magnetic field lines and will be addressed in future works.

\section{Conclusions and Discussion}
\label{discussion}

In this Letter, we explored the behavior of multidimensional pair-producing discharges within the polar zone of pulsar magnetospheres. Our main goal was to study the transverse coherence of these discharges, i.e., to understand what sets the smallest size of clusters of magnetic field lines that simultaneously possess an unscreened electric field necessary to trigger QED pair production. We found that current-driven discharges function in a cyclic pattern similar to the 1D model described by \cite{Timokhin2010, Timokhin2013}. Moreover, we demonstrated that the synchronization between localized discharges across magnetic field lines occurs during plasma-starved episodes when the unscreened electric field on different lines can communicate as it would happen in a vacuum. Our main conclusion is that the transverse coherence scale of a discharge zone is comparable to the longitudinal gap size, $l_{\rm gap}$. 

\subsection{Non-existence of sparks}

We studied the possibility of spark formation, namely the existence of long-lived isolated discharge columns surrounded by plasma-depleted regions where particle acceleration and pair formation are suppressed due to the accelerating electric field being screened by the presence of sparks. In a purely electrostatic problem, like in a discharge between the plates of a flat capacitor, the accelerating potential along the magnetic field lines between spark columns is limited by the potential across the magnetic field between the sparks, the latter, being filled with plasma at the latter stage of the discharge, acting essentially as conducting boundaries for inter-spark space \citep{Ruderman1975}. This would lead to the polar cap filled with sparks separated by the distance $\sim l_{\rm gap}$. However, the problem of polar cap discharges at the base of the FFE magnetosphere is not an electrostatic one, the magnetosphere requires a certain twist of magnetic field line and hence a certain current density $j_{{\rm mag}}$ to support it. If the actual current density deviates from $j_{{\rm mag}}$, a strong displacement current, $\partial{}E/\partial{}t$, appears -- the accelerating electric field is current-driven. Hence, a strong electric field should appear between discharge columns, regardless of how small the separation between them is, if the twist of magnetic field lines requires a current that can not be sustained by plasma without pair production.  So, the polar cap should be filled by discharges without empty zones if the current density supports pair formation. 
Our numerical simulations confirm this expectation.

In our setup, initially, we let a few discharge columns develop while keeping the accelerating electric field screened by dense plasma in the space between them. When these inter-discharge spaces are let to evolve freely, an accelerating electric field appears in them and drives pair formation. The whole polar cap becomes filled with discharges, with no plasma-depleted space left between them. This is true for both RS and SCLF regimes, as is evident in Figs.~\ref{fig:FullSCLF}-\ref{fig:FullRS}. If the width of the initial discharge column is smaller than the gap height, these columns expand until their width becomes comparable to the gap height; otherwise, a new discharge column will appear in the initial inter-discharge space. As in the 1D case, the new discharge is started by particles from the tail of the previous bursts of pair formation and not by particles coming from the NS surface, thus making the specific conditions for particle escape at the NS surface irrelevant for the start of the subsequent discharge.   Our code does not prevent the $\bm{E}\times\bm{B}$ drift, but we do not see any noticeable pair plasma drift in our simulations. This is to be expected as the accelerating potential variation across magnetic field lines is too low to make plasma move with respect to the NS with the velocity large enough to be visible at timescales of pair discharges \citep{vanLeeuwenTimokhin2012}. 

For the small gap size, $l_{\rm gap}\ll \PC{R}$, in the lava-lamp type cascades, the communication of discharges across the magnetic field lines results in the fragmentation of the polar cap into multiple densely packed coherent discharges with characteristic transverse size $\sim l_{\rm gap}$, with no plasma-depleted regions between them. Even if separate emission regions are directly associated with the plasma flows within individual discharges, the resulting emission pattern would not result in individual subpulses. On the timescale of a single pulse, which is much longer than the discharge repetition rate, each discharge would contribute comparable flux to the overall broad emission profile. In principle, one could think of a possibility for subpulse-like features to appear due to individual discharges if the emission is generated (or decouples from the plasma) at the boundaries of individual discharges. This can be possible for emission mechanisms recently proposed by \citet{Philippov2020, Melrose_Model2021}. The resulting emission pattern might look like a distorted ``honeycomb'', rather than a bunch of spots within the polar cap. However, even in the latter case, drifting subpulses seen in some pulsars can not be caused by individual discharge columns. Pulsars exhibiting subpulse drift phenomena are biased closer to the death line in the $P-\dot{P}$ diagram, which means that they have large gaps comparable to or exceeding the radius of the polar cap. We demonstrated that for these conditions, the discharges are proceeding coherently across magnetic field lines, ruling out the presence of many independent discharge zones in old-ish pulsars. Such pulsars can not have distinct emission zones associated with separate discharge columns. 

In our simulations, we start with the polar cap filled by plasma and find that in the well-developed phase, the discharges operate on all magnetic field lines where pair formation is possible, regardless of the conditions on particle supply from the NS surface. In other words, an existing FFE magnetospheric configuration can always support itself by generating plasma along all magnetic field lines, with no plasma-depleted regions between discharges. We conclude that a spark-filled polar cap is incompatible with the FFE pulsar magnetosphere.

\subsection{Implication for discharges in old pulsars}

In our paper, we did not consider gaps exceeding the size of the polar cap, as would be appropriate for old, less energetic pulsars. Their magnetospheres can differ from the fully FFE configurations assumed in this work. While in FFE solutions, the twist of the field lines is dictated by the conditions at the light cylinder, our local gap simulations show significant untwisting of the field lines that are unable to fully adjust to the magnetospheric current. This is particularly prominent in the initial phases of our simulations when extended thin patches are unable to ignite the ``current-driven'' discharge because of the voltage limitation caused by the transverse extent. Such discrepancies between the local current produced in the discharge zone and the global FFE magnetospheric solution will likely result in large-scale ``breathing'' of the magnetosphere (see Sect.~\ref{sec:curEvol}) on timescales comparable to the light-crossing time of the light-cylinder, $\sim P/2\pi$, with potential implications for the short-term radio nulling. Understanding these issues requires global simulations that include accurate modeling of pair production in the polar discharge for conditions appropriate for old pulsars.

\subsection{On repetition rate of the cascades}

A significant issue beyond this Letter is the repetition rate of the pair cascades -- the time interval between two successive bursts of pair formation along the same magnetic field lines, $\Delta\tau_{\rm{rep}}$. As we conjectured in Sec.~\ref{sec:SCLFtails}, it is likely that under realistic conditions, the small repetition rate (large $\Delta\tau_{\rm{rep}}$) is caused by a continuous flux of particles reversed from the pair cloud propagating away from the discharge zone. Significant improvements, even in 1D simulations, are required to accurately capture the long-term evolution of electric fields in pair clouds and resolve this issue. A particularly stunning example for the importance of the long-term evolution of discharges is the repetition rate on field lines carrying the return current. In local simulations, the discharge starts at the upper boundary of the simulation domain, however, it is unclear how far the plasma-starved zone extends into the magnetosphere. It is possible that the region with the unscreened electric field can first form at the null point, where the GJ charge density is zero \cite[e.g.,][]{Bransgrove2023}. If this assumption holds, it is likely that electric fields could easily synchronize across the extended return current zone, significantly enhancing the coherence of the discharge. This is another question that should be addressed using next-generation global simulations that include an accurate discharge model. 

\section{acknowledgments}

We thank Hayk Hakobyan, Anatoly Spitkovsky, and Chris Thompson for fruitful discussions and an anonymous referee for useful comments. This work was supported by NSF Grant No. PHY-2231698, and facilitated by Multimessenger Plasma Physics Center (MPPC), NSF Grant No. PHY-2206607. This work was supported by a grant from the Simons Foundation (00001470, AP). AP acknowledges support by an Alfred P. Sloan Fellowship. AT was supported by the grant 2019/35/B/ST9/03013 of the Polish
National Science Centre.  Computing resources were provided and supported by the Division of Information Technology at the University of Maryland (\href{http://hpcc.umd.edu}{\texttt{Zaratan} cluster}\footnote{\url{http://hpcc.umd.edu}}). This research is part of the Frontera computing project \citep{Stanzione2020} at the Texas Advanced Computing Center (LRAC-AST21006). Frontera is made possible by NSF award OAC-1818253.

\appendix

\section{Additional Details on radiation and QED physics}
\label{rescalingQED}

The radiation reaction force of the curvature radiation and the corresponding energy losses are given by
\begin{equation}
    m c \frac{d{\bm u}}{d t} = - \frac{2}{3} e^2 \frac{\gamma^4}{\rho_{\rm c}^2} {\bm \beta},~~~\frac{d E}{d t} \equiv -I = - \frac{2}{3} e^2 c \frac{\gamma^4}{\rho_{\rm c}^2} \beta^2, 
    \label{RadLosses}
\end{equation}
where $\rho_{\rm c}$ is the curvature radius of the field line. The emitted photons have a peak energy of 
\begin{equation}
    \varepsilon_{\rm ph}^* = \frac{3}{2}\hbar \frac{c}{\rho_{\rm c}} \gamma^3,
    \label{PhotonEnergy1}
\end{equation}
and the spectral emissivity is given by the synchrotron kernel \citep{Rybicki1979}
\begin{equation}
    \frac{d I(\varepsilon)}{d \varepsilon} = A \frac{\varepsilon}{\varepsilon_{\rm ph}^*} \int_{\frac{\varepsilon}{\varepsilon_{\rm ph}^*}}^\infty K_{5/3}(x)dx
    \label{K53}
\end{equation}
where $K_{5/3}(x)$ is the Macdonald function, and $\varepsilon_{\rm ph}^*$ is given by Equation (\ref{PhotonEnergy1}). The coefficient A is determined by equating the full emitted power, Eqn. (\ref{RadLosses}), with the losses integrated over the frequency spectrum. For the normalized photon energy, $\Tilde{\varepsilon}=\varepsilon/m_e c^2$, the photon emission rate is 
\begin{equation}
    \frac{d N}{d t d \Tilde{\varepsilon}} =  \frac{1}{\sqrt{3}\pi} \frac{e^2 m_e c^2}{\hbar^2 c} \frac{1}{\gamma^2} \int_{\frac{\Tilde{\varepsilon}}{\Tilde{\varepsilon}_{\rm ph}^*}}^\infty K_{5/3}(x) dx = \frac{1}{\sqrt{3}\pi} \frac{\fine{\alpha} c}{\lambda_c}  \frac{1}{\gamma^2} \int_{\frac{\Tilde{\varepsilon}}{\Tilde{\varepsilon}_{\rm ph}^*}}^\infty K_{5/3}(x) dx,
    \label{PhotonEmission}
\end{equation}
where $\fine{\alpha}=e^2/\hbar c$ is the fine structure constant, and $\lambda_c = \hbar/m_e c$ is the reduced Compton wavelength.

To rescale the efficiency of curvature cooling, we choose the Lorentz factor of a particle, $\gamma_{\rm rad}$, for which curvature radiation reaction force balances acceleration in a vacuum electric field on a scale of the polar cap, $\PC{E}$:
\begin{equation}
    e \PC{E} = \frac{2}{3} e^2 \frac{\gamma_{\rm rad}^4}{\rho_{{\rm c}, 0}^2},~~~\PC{E} = 4 \pi \GJ{\rho} \PC{R}
\end{equation}
where $\PC{R}$ is the radius of the polar cap; and the vacuum electric field can be found as:
\begin{equation}
    \nabla \cdot {\bm E}\equiv \frac{d E_z}{d z}= -4 \pi e \GJ{n} \Longrightarrow E_z(z) = -4 \pi e \GJ{n} z.
\end{equation} 
A convenient way to parameterize the size of the polar cap is to express it through the Lorentz factor reached by a particle in the full potential drop:
\begin{equation}
    m_e c^2 \PC{\gamma} =4 \pi e \int_0 ^{\PC{R}} e \GJ{n} z dz = 4 \pi e^2 \GJ{n} \PC{R}^2/2 \longrightarrow  \PC{R}^2 = \frac{m_e c^2}{4\pi \GJ{n} e^2} 2 \PC{\gamma} \equiv \left(\UGJ{d_{\rm e}}\sqrt{2 \PC{\gamma}}\right)^2.
\end{equation}
Curvature radius of the field line, $\rho_{{\rm c}, 0}$, can be defined as: 
\begin{equation}
    \rho_{{\rm c}, 0} = \PC{R} \Theta = \UGJ{d_{\rm e}} \sqrt{2 \PC{\gamma}} \Theta,~~~~\Theta \gg 1.
\end{equation}

Another energy scale defines the Lorentz factor of a particle producing a photon of energy equal to the electron rest mass:
\begin{equation}
    m_e c^2 \equiv \frac{3}{2}\hbar \frac{c}{\rho_{{\rm c}, 0}} \gamma_{\rm emit}^3. 
\end{equation}

These equations allow to parametrize the energies of curvature photons and the strength of the radiation reaction force. To summarize,
\begin{equation}
    \PC{R}= \UGJ{d_{\rm e}} \sqrt{2 \PC{\gamma}}, ~~~ \rho_{{\rm c}, 0} = \PC{R} \Theta, ~~~F_{\rm curv} =  e^2 \GJ{n} \PC{R} \left(\frac{\gamma}{\gamma_{\rm rad}}\right)^4 \left(\frac{\rho_{\rm c}}{\rho_{{\rm c}, 0}}\right)^{-2}, ~~~\Tilde{\varepsilon}_{\rm ph}^* = \frac{\rho_{{\rm c}, 0}}{\rho_{\rm c}} \left(\frac{\gamma}{\gamma_{\rm emit}} \right)^3
\end{equation}

With these renormalizations, the spectral photon emission rate (\ref{PhotonEmission}) is
\begin{eqnarray}
    \frac{d N}{d t  d \Tilde{\varepsilon}} &=& \frac{9\sqrt{3}}{8\pi} \frac{1}{\varepsilon_{\rm ph}^*} I \frac{1}{\Tilde{\varepsilon}_{\rm ph}^*} \int_{\frac{\Tilde{\varepsilon}}{\Tilde{\varepsilon}_{\rm ph}^*}}^\infty K_{5/3}(x) dx= \frac{9\sqrt{3}}{8\pi} c \frac{e^2  \GJ{n} \PC{R}}{m_e c^2}\left(\frac{\gamma}{\gamma_{\rm rad}}\right)^4 \left(\frac{\rho_{\rm c}}{\rho_{{\rm c},0}}\right)^{-2} \frac{1}{\left(\frac{\gamma}{\gamma_{\rm emit}}\right)^6 \left(\frac{\rho_{\rm c}}{\rho_{{\rm c}, 0}}\right)^{-2}} \int_{\frac{\Tilde{\varepsilon}}{\Tilde{\varepsilon}_{\rm ph}^*}}^\infty K_{5/3}(x) dx = \nonumber \\
&=&\frac{9\sqrt{3}}{4\sqrt{2}\pi} \frac{c \PC{R}}{(\UGJ{d_{\rm e}})^2} \frac{\gamma_{\rm emit}^6}{\gamma_{\rm rad}^4} \frac{1}{\gamma^2} \int_{\frac{\Tilde{\varepsilon}}{\Tilde{\varepsilon}_{\rm ph}^*}}^\infty K_{5/3}(x) dx = \frac{9\sqrt{3}}{4\sqrt{2}\pi} \frac{c}{\UGJ{d_{\rm e}}} \frac{\PC{\gamma}^{1/2} \gamma_{\rm emit}^6}{\gamma_{\rm rad}^4} \frac{1}{\gamma^2} \int_{\frac{\Tilde{\varepsilon}}{\Tilde{\varepsilon}_{\rm ph}^*}}^\infty K_{5/3}(x) dx 
\label{RenormPhotonEmission}
\end{eqnarray}
In the code, we employ tabulated values of $T(y_j)=\int_{y_j}^\infty K_{5/3}(x) dx$ with $y=\Tilde{\varepsilon}/\Tilde{\varepsilon}_{\rm rad}^*$.

\begin{figure*}[t]
\includegraphics[width=\textwidth]{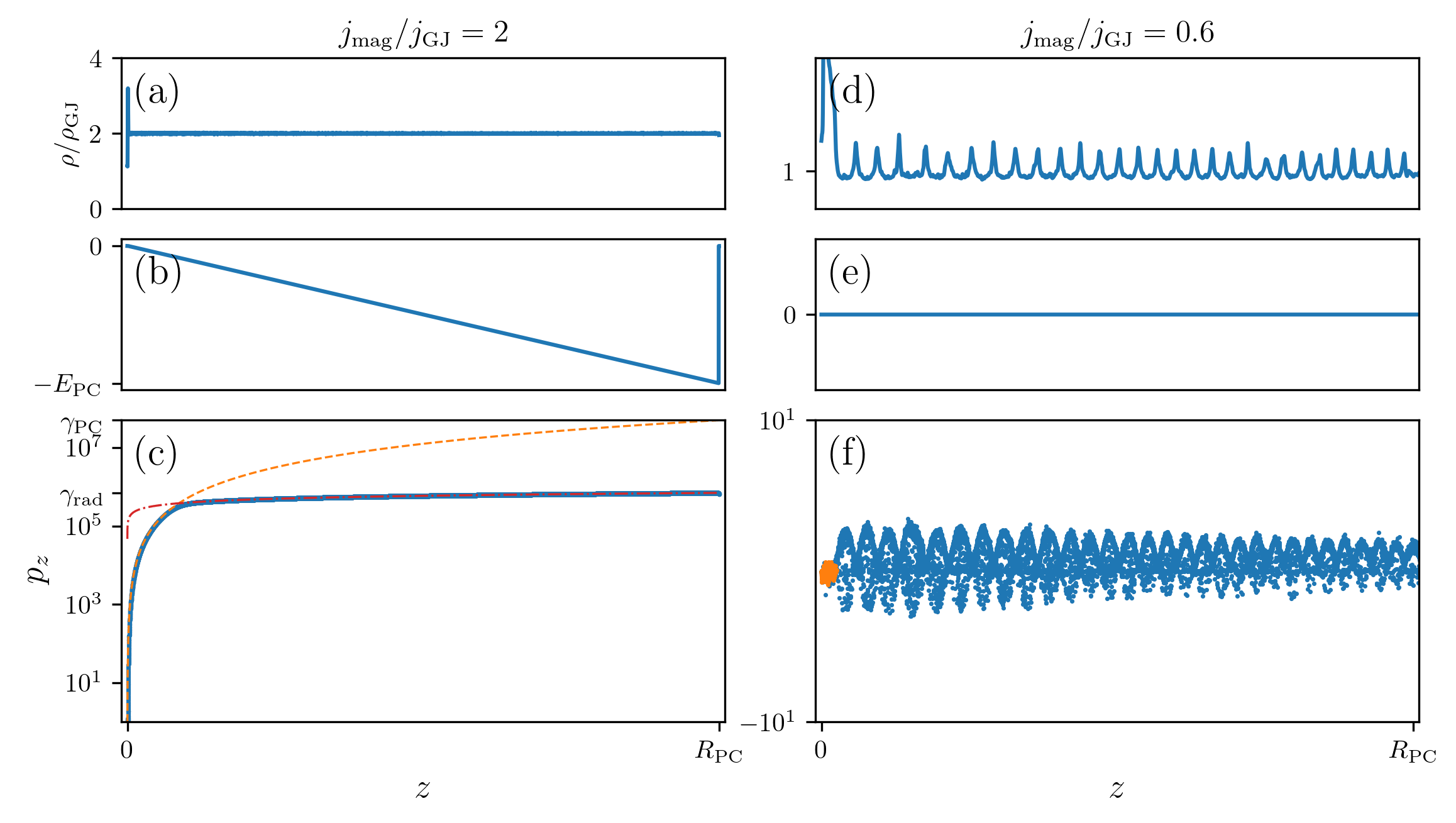}
\caption{1D test of the curvature cooling and atmospheric particle injection in the charge-separated SCLF model:  the left column shows (a) plasma density, (b) electric field, and (c) phase space, $z-p_z$ for the super-GJ current; and (d-g) - same quantities for the sub-GJ current. For the super-GJ current, the phase space plot, (c), shows that electrons initially accelerate in almost vacuum potential (the orange dotted line shows analytical prediction) and later enter into the radiation reaction limited regime (indicated by the red dot-dashed line). In the case of the sub-GJ current, extracted electrons travel at mildly relativistic speeds and maintain a state of small electric field.}
\label{fig:testCooling}
\end{figure*}

The differential cross section of the pair production reaction $\gamma+B \rightarrow e^- + e^+$ is given by \citep{Erber1966}:
\begin{equation}
    d\sigma = 0.23 \frac{B}{B_q} \sin{\psi} \frac{\alpha_F}{\lambda_c} \exp{\left(-\frac{8}{3 \frac{B}{B_q} \sin{\psi} \Tilde{\varepsilon}_{\rm ph}}\right)} dl.
\end{equation}
Here, $B_q = m_e^2 c^3/e\hbar \approx 4.41 \times 10^{13}$G is the critical magnetic field, and $\psi$ is the angle between the magnetic field and the direction of photon propagation. Since a photon is emitted along the field line and accumulates angle during the propagation, it can be estimated as $\psi(x) = \int_0^z dz/\rho_{\rm c}(z)$, where $\rho_{\rm c}(z)$ is the curvature of the local field line. To renormalize the cross-section in the simulation, we need to express $\fine{\alpha}/\lambda_c$ in the code units. Direct comparison of Equations (\ref{PhotonEmission}) and (\ref{RenormPhotonEmission}) gives:
\begin{equation}
    \frac{1}{\sqrt{3}\pi} \frac{\fine{\alpha} c}{\lambda_c}  \frac{1}{\gamma^2} \int_{\frac{\Tilde{\varepsilon}}{\Tilde{\varepsilon}_{\rm ph}^*}}^\infty K_{5/3}(x) dx = \frac{9\sqrt{3}}{4\sqrt{2}\pi} \frac{c}{\UGJ{d_{\rm e}}}\frac{\PC{\gamma}^{1/2} \gamma_{\rm emit}^6}{\gamma_{\rm rad}^4} \frac{1}{\gamma^2} \int_{\frac{\Tilde{\varepsilon}}{\Tilde{\varepsilon}_{\rm ph}^*}}^\infty K_{5/3}(x) dx \Longrightarrow \frac{\fine{\alpha}}{\lambda_c} = \frac{27}{4\sqrt{2}}\frac{1}{\UGJ{d_{\rm e}}} \frac{\PC{\gamma}^{1/2} \gamma_{\rm emit}^6}{\gamma_{\rm rad}^4}.
\end{equation}
Then, the resulting differential cross-section is 
\begin{equation}
     d\sigma = 0.23 \frac{27}{4\sqrt{2}} \frac{B}{B_q} \frac{1}{\UGJ{d_{\rm e}}} \frac{\PC{\gamma}^{1/2} \gamma_{\rm emit}^6}{\gamma_{\rm rad}^4} \sin{\psi} \exp{\left(-\frac{8}{3 \frac{B}{B_q} \sin{\psi} \Tilde{\varepsilon}_{\rm ph}}\right)} dl.
     \label{dSigmaNew}
\end{equation}
We compute the total optical depth, $\sigma(l)=\int_0^l d\sigma$, during the propagation of a photon, as the angle between the magnetic field and the photon wave vector increases, $d\psi=c \Delta t/\rho_{\rm c}(z)$, at every time step. When the optical thickness reaches unity, we simulate the photon decay into an electron-positron pair by removing the photon and adding an electron-positron pair at the same location, with the 4-velocity along the field line
\begin{equation}
    u_{||} = \frac{|\cos{\psi}|(\Tilde{\varepsilon}_{\rm ph}^2-4)^{1/2}}{\left(\Tilde{\varepsilon}_{\rm ph}^2 \sin^2{\psi}+4\cos^2{\psi}\right)^{1/2}} \frac{(\textbf{k}\cdot\textbf{b})}{|\textbf{k}||\textbf{b}|},
\end{equation}
where $\textbf{k}$ and $\textbf{b}$ are unit vectors in the directions of the photon wave vector and the magnetic field, correspondingly.

In order to test both the plasma supply from the stellar atmosphere and the efficiency of curvature cooling, we employ a 1D charge-separated setup with different values of the magnetospheric currents, following \citep{Timokhin2013}. In Fig.~\ref{fig:testCooling}, we show tests for two different magnetospheric currents: sub-GJ, $j_{\rm mag}=0.6 \GJ{j}$, and super-GJ, $j_{\rm mag}=2 \GJ{j}$. In the case of the sub-GJ current, the magnetospheric current extracts the charge-separated flow of electrons with density $\approx \GJ{\rho}$, moving with sub-relativistic velocity (subplots (d) and (f), respectively), which leads to an efficient screening of the electric field (subplot (e) of Fig. \ref{fig:testCooling}). In the case of the super-GJ current, the larger electron density is extracted, $n\approx |j_{\rm mag}|/(ec)$, which leads to a strong electric field, $E_{z} \approx 4\pi(j_{\rm mag}/c-\GJ{\rho})z$ (see subplots (a) and (b) in Fig. \ref{fig:testCooling}) and quick electron acceleration. During the initial phases of acceleration, the cooling to curvature radiation is negligible, and the acceleration is determined by the electric field, $\gamma(z)=2\pi |e \GJ{\rho}| z^2/ m_e c^2$ (orange dashed line in Fig. \ref{fig:testCooling}c). When the Lorentz factor becomes comparable to $\gamma_{\rm rad}$, the electron enters a radiation-reaction-limited regime when the radiation losses are balanced by an acceleration in the electric field $\gamma=\gamma_{\rm rad}(E_z(z)/\PC{E})^{1/4}=\gamma_{\rm rad}(z/\PC{R})^{1/4}$ (shown as red dashed-dotted line). We also verified that the total energy of the emitted photons is equal to the full radiative cooling losses experienced by the particle.  

\bibliography{sample631}{}
\bibliographystyle{aasjournal}


\end{document}